\documentclass[%
 reprint,
amsmath,amssymb, aps,
]{revtex4-2}

\usepackage[english]{babel}
\usepackage[utf8]{inputenc}
\usepackage[colorinlistoftodos, color=blue!40, prependcaption]{todonotes}
\usepackage{orcidlink}

\usepackage{amsthm}
\usepackage{mathtools}
\usepackage{physics}
\usepackage{xcolor}
\usepackage{graphicx}
\usepackage[left=17mm,right=17mm,top=20mm,columnsep=15pt]{geometry} 
\usepackage{adjustbox}
\usepackage{placeins}
\usepackage[T1]{fontenc}
\usepackage{lipsum}
\usepackage{csquotes}
\usepackage[percent]{overpic}
\usepackage{multirow}

\usepackage{bm}

\usepackage{hyperref}
\hypersetup{
    colorlinks=true,
    linkcolor=blue,
    citecolor = blue,
    filecolor=magenta,      
    urlcolor=blue,
    pdftitle={Overleaf Example},
    pdfpagemode=FullScreen,
    }

\begin{document}

\title{Radial velocity statistics of cosmic voids as a probe of interacting dark energy}

\newcommand{\inst}[1]{\textsuperscript{#1}}

\author{
Kin Ho Luo\inst{1} \orcidlink{0009-0009-1698-700X}}
\email{ 1155129240@link.cuhk.edu.hk}
\author{Ming-chung Chu\inst{1} \orcidlink{0000-0002-1971-0403}}
\author{Kwan Chuen Chan\inst{2} \orcidlink{0000-0001-8757-408X}}
\author{Wangzheng Zhang\inst{3} \orcidlink{0000-0003-0102-1543}}

\affiliation{
\inst{1} Department of Physics, The Chinese University of Hong Kong, Sha Tin, N.T., Hong Kong\\
\inst{2} School of Physics and Astronomy, Sun Yat-sen University, 2 Daxue Road, Tangjia, Zhuhai, 519082, China \\
\inst{3} Institut d'Astrophysique de Paris, UMR~7095, CNRS, Sorbonne Universit\'e, 98~bis~boulevard Arago, 75014~Paris, France 
}

\date{\today} 

\begin{abstract}

Due to their vast sizes and extremely low densities, the dynamics of cosmic voids are largely decoupled from complex, small-scale baryonic physics and are highly sensitive to the background expansion of the Universe. This makes them clean and sensitive probes of dark energy's properties. Using N-body simulations, we show that the void radial velocity and velocity dispersion profiles are sensitive to the Type 3 interacting dark energy model parameters, the momentum coupling $\beta$ ($<0$) and the scalar field parameter $\lambda$. Within the $1\sigma$ range of the best-fit values of $\beta$ and $\lambda$ constrained by Planck CMB, DESI BAO, and DES-Y5 supernova data, we find up to $\sim 30\%$ deviations in the void radial-velocity and velocity-dispersion profile spans relative to the uncoupled scenario, which can be well-approximated by a 4-parameter quadratic regression model. This demonstrates that the void radial velocity statistics provide an independent and observationally accessible probe of the dark-sector interaction in the Type 3 model. 



\end{abstract}


\maketitle

\section{Introduction} \label{sec:introduction}

In the standard paradigm of cosmological structure formation, gravity amplified primordial fluctuations, giving rise to the large-scale structure known as the cosmic web, a complex arrangement of sheets and filaments that intersect at dense clusters \cite{Zeldovich1970}. In contrast, regions originating from primordial underdensities possessed a lower average density than the cosmic mean. Experiencing a weaker inward gravitational pull compared to the background Universe, the expansion of these regions was less hindered by gravity than the cosmic mean. As a result, they expanded more rapidly than the Hubble flow, forming vast underdense areas referred to as cosmic voids \cite{Gregory1978}.

In recent years, voids have attracted significant scientific interest. As the most underdense regions in the Universe, they occupy a substantial fraction of cosmic volume, making them powerful probes of cosmology and structure formation. For instance, because the large-scale bias of voids is highly sensitive to the initial conditions of the density field, their clustering provides a powerful tool to constrain primordial non-Gaussianity \cite{Chan_etal2019}. Additionally, since their typical size is comparable to the neutrino free-streaming scale \cite{Kreisch2019}, the threshold above which neutrinos cluster, voids are particularly useful for constraining neutrino properties \cite{Massara2015, Banerjee2016, Kreisch2019, Schuster2019, Zhang2020, Contarini2021, Bayer2021, Kreisch2022}. Specifically, neutrino free-streaming leads to a scale-dependent reduction in the growth of matter fluctuations, affecting the formation of substructure within void regions and modifying void density and velocity profiles relative to those for $\Lambda$CDM with massless neutrinos. Furthermore, their extremely low internal densities and shallow gravitational potentials provide an ideal environment to test modifications of general relativity (GR) \cite{Zivick2015, Barreira2015, Falck2018, Baker2018, Paillas2019, Davies2019, Perico2019, Tamosiunas2022, Fiorini2022}. This is because in many of these models, the screening mechanism is inefficient in low-density environments. As a result, the unscreened fifth force inside the void provides an enhanced pull directed radially outward towards the higher-density compensation wall. This enhanced effective gravity accelerates the evacuation of matter from the void interior to the compensation wall. These features also allow voids to be uniquely sensitive to diffuse components in the Universe, such as the dark energy (DE) \cite{Lee2009, Bos2012, Lavaux2012, Sutter2015b, Pisani2015, Pollina2016, Verza2019, Verza2023}.

Various void properties have emerged as promising probes of DE. The temporal evolution of void ellipticities for $\Lambda$CDM and quintessence models has been shown to be different and can be used to constrain the DE equation of state (EOS)  $\omega_{\text{DE}} (a)$, with $a$ being the cosmological scale factor \cite{Lee2009, Bos2012}.  Void abundance offers another robust constraint: \cite{Pisani2015} demonstrated that accounting for the number of voids as a function of redshift in forecasts for \textit{Euclid} \cite{Hamaus2022, Scaramella2022, Cropper2025} and \textit{Roman} \cite{Spergel2015} significantly improves constraints on the CPL parametrization of $\omega_{\text{DE}}(a)$, $\omega_{CPL}(a) = \omega_0 + \omega_a(1-a)$, with $\omega_0$ and $\omega_a$ being constants,  compared to traditional probes without void counts. Additionally, \cite{Sutter2015b} and \cite{Pollina2016} explored stacked density profiles and size distributions of voids, and showed that the fifth force in coupled dark energy (cDE) models empties voids, resulting in larger and more underdense voids than in $\Lambda$CDM.

Voids exhibit universal characteristics that make them valuable cosmological probes. One such property is their radial density profile, which is well described by a universal functional form \cite{Hamaus2014} and quantifies structure formation within and around voids. Crucially, local mass conservation allows this profile to be related to the void radial velocity profile $v_\text{v}(r)$ via the continuity equation. $v_{\text{v}}(r)$ plays a central role in interpreting redshift-space distortions (RSD)—distortions in observed clustering caused by the peculiar velocities of tracers along the line of sight—and the Alcock–Paczynski (AP) effect, a geometric distortion that arises from assuming the wrong cosmology (expansion history and cosmological parameters) when converting observed redshifts and angular separations to comoving coordinates, leading to apparent anisotropy in intrinsically isotropic structures. 

Beyond its established role as an estimator of RSD, $v_\text{v}(r)$ serves as a sensitive probe of the large-scale structure (LSS). $v_\text{v}(r)$ has been widely used to constrain cosmological parameters such as $\Omega_m$ \cite{Hamaus2016}, the linear growth rate \cite{Hamaus2015, Hamaus2016, Hawken2017, Nadathur2019, Correa2022, Massara2022}, neutrino masses, and modified gravity theories including $f(R)$ and nDGP \cite{Cai2015, Wilson2023}. Moreover, it has been employed to investigate void evolution and dynamics, distinguishing expanding large voids from collapsing small ones \cite{Paz2013, Ceccarelli2013, Massara2018}, and to probe the AP effect for geometric cosmology tests \cite{Hamaus2015}.

Similarly, the void radial velocity dispersion $\sigma_\text{v}(r)$  has been used to constrain modified gravity models such as $f(R)$ and nDGP \cite{Cai2015, Wilson2023}, study tracer bias and void evolution in low-density environments \cite{Nadathur2019}, and improve RSD modeling for growth rate measurements \cite{Hamaus2015, Cai2015, Hamaus2016, Correa2022}. More recently, it has been employed to refine linear-theory predictions for void-galaxy correlation functions, highlighting environmental dependencies in the velocity fields \cite{Correa2022, Massara2022}. Together, $v_\text{v}(r)$ and $\sigma_\text{v}(r)$ provide a promising framework for testing cosmological models using the unique dynamical environment of cosmic voids.


This potential is particularly relevant in the context of dynamical DE models, which have gained renewed attention following evidence from DESI DR2 suggesting a time-varying DE EOS \cite{Adame2025a, Adame2025b, Adame2025c}, which also suggests the possibility of nongravitational interaction between the dark components, such as in the interacting dark-energy (IDE) models.  

The most popular IDE models involve energy-momentum interaction between DE and dark matter (DM) \cite{amendola2000, di_valentino2020b, di_valentino2021}. These models offer a  natural explanation of the cosmic coincidence such that $\Omega_{\Lambda, 0}/\Omega_{c, 0} \sim \mathcal{O}(1)$, in which $\Omega_{\Lambda, 0}$ and $\Omega_{c, 0}$ denote the energy densities of DE and DM at $z=0$, respectively. The relevant mechanisms seem to offer alleviation of the Hubble tension \cite{di_valentino2017, di_valentino2021b}. For example, by considering a phantom-like DE EOS ($\omega_\mathrm{DE}(a)<-1$) in which DM decays into DE, \cite{di_valentino2017} reports an increase in the constrained Hubble constant ($H_0$) value relative to that of the $\Lambda$CDM, easing the $H_0$ tension. However, these models exhibit noticeable shortcomings, such as large quantum corrections that can exacerbate the cosmological constant problem \cite{damico2016, marsh2017} and the reliance on ad hoc, purely phenomenological coupling parameters that fit the CMB data worse than $\Lambda$CDM's \cite{GomezValent2020, Bean2008, Xia2009}. In most IDE models, there is an energy exchange between the two dark components, which would modify the background evolution and lead to nonadiabatic instability in dark sector perturbations \cite{valiviita2008}.

Recently, a new class of pure momentum transfer models has received special attention: the Type 3 model \cite{pourtsidou2013}. This model addresses the shortcomings of IDE models by considering momentum transfer within the dark sector only \cite{pourtsidou2013}, parameterizing the momentum exchange with a coupling constant $\beta$ and the DE EOS with the slope of scalar field potential $\lambda$. For certain combinations of $\beta$ and $\lambda$, this model leaves the background expansion history unchanged with respect to that of the $\Lambda$CDM model, achieving an excellent fit with the CMB data \cite{pourtsidou2016}. With a negative $\beta$, it has even been shown to alleviate the $S_8$ tension \cite{pourtsidou2016, chamings2020, Pourtsidou2026}. Furthermore, the Type 3 model introduces DE-DM couplings at the level of the action, providing a more intuitive and direct insight into the interactions and their impact on the background expansion and structure formation. Because this model is described by a Lagrangian formalism, instabilities can be easily determined and avoided \cite{chamings2020}. 

Given the established sensitivity of both matter pairwise velocity statistics \cite{Luo2026} and various void properties to DE, it is highly plausible that void radial velocity statistics is also sensitive to the distinctive signatures of IDE models, offering a new avenue for testing these scenarios.

Building on our previous study of matter pairwise velocities in the Type 3 model \cite{Luo2026}, this paper investigates the dependence of $v_{\text{v}}(r)$ and  $\sigma_{\text{v}}(r)$ on the two Type 3 model parameters, $\beta$ and $\lambda$. 
The organization of this paper is as follows. Section \ref{sec: theory and sim} introduces the theoretical formalism of the Type 3 model and its implementation in N-body simulations. Section \ref{sec:velocity} defines $v_{\text{v}}(r)$ and $\sigma_{\text{v}}(r)$ and discusses their physical interpretation. Section \ref{sec:results} presents the results, and Section \ref{sec:conclusions} summarizes and concludes the paper.

\section{Model and simulations} 
\label{sec: theory and sim}
\subsection{The Type 3 model}
The defining feature of the Type 3 model constructed in \cite{pourtsidou2013} is that the energy transfer between DE and DM, embodied by the coupling current $ J_{\mu} $, is absent at the background level in the energy conservation equations. This ensures no energy exchange between the dark components at either the background or the linear perturbation level across all values of $\beta < 1/2$ and $\lambda$. The momentum coupling constant is strictly restricted to $\beta<1/2$. Beyond this value, the scalar field develops a negative kinetic energy term, leading to an unphysical 'ghost' instability \cite{pourtsidou2013}.


The Type 3 model is described by the Lagrangian
\begin{equation} L = F(Y,Z,\phi) + g(n_c) , \label{ori_lag} \end{equation}
where $ \phi $ is the scalar field describing DE, $ Y = \frac{1}{2} (\nabla_{\mu} \phi)^2 $ is the scalar field's kinetic energy density. Additionally, $ n_c $ is the number density of DM particles, and $ Z = u^{\mu} \nabla_{\mu} \phi $ encodes the interaction between the DM fluid's four-velocity $ u^{\mu} $ and the scalar field's gradient, enabling momentum exchange within the dark sector.  Here, $F$ is the scalar field Lagrangian governing the DE dynamics, while $g$ is the uncoupled DM fluid Lagrangian.

Based on Eq.~\eqref{ori_lag}, we adopt a quintessence-like scalar field with a constant sound speed $ c_s = 1 $, incorporating a quadratic coupling term. This yields the scalar field Lagrangian 
\begin{equation} F = Y + \beta Z^{2} + V(\phi). \label{scf_lag} \end{equation}

Here, $\beta$ is the momentum coupling constant, and the scalar field potential is $V(\phi)=V_{0}e^{-\lambda\kappa\phi}$, where $ \kappa = m_P^{-1} = \sqrt{8\pi G} $ is the inverse of the reduced Planck mass. We work in natural units ($c=\hbar =1$) where the scalar field $ \phi $ has dimensions of mass (expressed in units of $ m_P $), and the potential takes a form inspired by inflationary theory. Overall, $\beta$ and $\lambda$ modify the kinetic and potential energy of the scalar field, respectively.

In Eq.~\eqref{scf_lag}, the additional term $\beta Z^2$ in the scalar field Lagrangian 
represents the interaction between the dark components. As demonstrated by the linear perturbation analysis in \cite{Luo2026}, for $\beta < 0$ this coupling results in a net transfer of momentum from the DM particles to the scalar field, introducing a frictional force that slows DM motion compared to the uncoupled case. This effect becomes more pronounced as the scalar field potential steepens because a larger $\lambda$ speeds up the field evolution ($\dot\phi$ increases), enhancing the kinetic coupling and momentum interaction between the dark components.


Additionally, from Eq.~\eqref{scf_lag}, we obtain the background Klein-Gordon equation governing the evolution of the scalar field \cite{pourtsidou2016}, 
\begin{equation}
\ddot{\phi} + 2{\cal{H}\dot{\phi}} + \left( \frac{1}{1 - 2\beta} \right) a^2 \frac{dV}{d\phi} = 0, 
\label{eq: scf expli equation}
\end{equation}
where the dot denotes the derivative with respect to conformal time $\eta$ and $\cal{H}$ is the conformal Hubble parameter.

Likewise, the background continuity equation for the DM fluid takes the form
\begin{equation}
    \dot{\rho}_{c} + 3{\cal{H}}{\rho_{c}} = 0. \label{evolution}
\end{equation} 

Here, $\rho_{c}=\rho_{c,0}a^{-3}$ is the DM fluid energy density, with $ \rho_{c,0} $ being its value at $ a = 1 $.

To examine how the Type 3 model impacts LSS, we analyze linear perturbations around the flat Friedmann-Lemaître-Robertson-Walker (FLRW) metric in the Newtonian gauge for the ease of physical interpretation and implementation in our N-body simulations. The perturbed metric is \cite{baumann2022}
\begin{equation}ds^2 = a^2(\eta) \bigl( -\left( 1+ 2\Psi \right) d\eta^2 + \left( 1- 2\Phi  \right)  dx^i dx_i \bigl),  \end{equation}
\\
where  $i = 1, 2, 3$ denote the spatial index and $\Psi$ and $\Phi$ are the scalar perturbation modes. Specifically, $\Psi$ corresponds to the Newtonian gravitational potential, while $\Phi$ represents the spatial curvature perturbation (Assuming the DM fluid has zero anisotropic stress, $\Psi = \Phi$).

Using Eq.~\eqref{evolution}, the evolution of the DM fluid perturbation is \cite{baumann2022} 
\begin{equation}
    \dot{\delta}_{c}= -\theta_{c} + 3\dot{\Phi} \label{perturbations},
\end{equation}
where $ \delta_c = \frac{\rho_c - \bar{\rho}_c}{\bar{\rho}_c} $ is the density contrast, $ \theta_c = \nabla_i v^i $ is the velocity divergence, and $ v^i $ denotes the perturbed three-velocity of the DM fluid (with $ i = 1, 2, 3 $). Both Eq.~\eqref{evolution} and Eq.~\eqref{perturbations} stem from treating DM as a pressure-less perfect fluid, and its mean density evolution remains unaltered by the momentum coupling.

On the other hand, the Euler equation is modified by the momentum coupling  \cite{Luo2026}, 
\begin{eqnarray} 
  \dot{v}_i + \gamma_1{\cal{H}} v_i + \gamma_2 \partial_i \Psi 
 =& 0, \label{eq:euler eq} 
\end{eqnarray}
with 
\begin{eqnarray}
\label{coefficients}
\nonumber c_3 & = & \frac{2\beta\dot{\phi}^2}{a^2 \rho_c - 2 \beta \dot{\phi}^2}, \\
\nonumber c_1 & = & 2\beta c_3, \\
 c_2 & = & 
\nonumber\left[ \left(3- 4\beta\right) + 2 a^2  \frac{dV}{d\phi} \frac{1}{\dot{\phi}} \frac{1}{\cal{H} } \right]c_3,\\
\nonumber  \gamma_1 & = & \frac{1+c_2}{1+c_1}, \\
\gamma_2 & = & \frac{1+c_3}{1+c_1}. \label{eq: coefficients}
\end{eqnarray}
Clearly, the coupling parameter $ \beta $ modulates the strengths of cosmological friction and gravitational force. In Eq.~\eqref{eq: coefficients}, because the $1^{\text{st}}$ term of the denominator of $c_3$, $a^2 \rho_c$, is always positive, a large deviation from the uncoupled case ($c_{j} = 0,\,\, j = 1, 2, 3$) is expected when $a^2\rho_{c}  \approx 2\beta\dot{\phi}^2$ for $\beta>0$ \cite{Luo2026}. 


\label{sec:develop}

\subsection{Simulations and Halo Catalogs}

The main feature of the Type 3 model, a quintessence scalar field with momentum coupling, is encapsulated by $\beta$ and $\lambda$. Therefore, we focus on these two parameters to probe the relationship between the momentum transfer in the dark sector and the void radial velocity statistics. 

To study how these parameters affect the cosmology at the linear level, we performed a Markov Chain Monte Carlo (MCMC) analysis using \texttt{Cobaya} \cite{Torrado2021}. In \cite{Pourtsidou2026}, MCMC constraints on $\beta$ and $\lambda$ were obtained by refitting the Type 3 model with Planck cosmic microwave background (CMB), lensing \cite{planck2020b, Rosenberg2022, Carron2022, Aghanim2019}, DESI DR2 baryon acoustic oscillations (BAO) \cite{AbdulKarim2025}, and DES-Y5 Type Ia supernova data \cite{DESCollaboration2024}. Using the same datasets, we ran additional MCMC chains for different combinations of $\beta$ and $\lambda$. In this paper, we consider negative values of $\beta$ only, since $\beta > 0$ yields a highly suppressed $H_0$ that is phenomenologically disfavored \cite{Pourtsidou2026} (see Appendix \ref{sec:mcmc_constraints}). Our chains converge when the Gelman-Rubin diagnostic \cite{Gelman1992} $R-1 \leq 0.01$, and they are analyzed using \texttt{GetDist} \cite{Lewis2025}. This analysis is essential for the N-body simulation, as the cosmological parameters govern the initial conditions, the expansion history, and the clustering of matter.

N-body simulations for the Type 3 model are subsequently performed using the modified Euler equation (Eq.~\eqref{eq:euler eq}) and the methodology outlined in \cite{Luo2026}. In summary, we adopt a modified version of \texttt{Gadget-2} \cite{Springel2005}, referred to as \texttt{ME-Gadget-2} \cite{An2019a, An2019b, zhang_gadget_2018}, to study 10 combinations of $\beta$ and $\lambda$ within their allowed ranges as presented in Table \ref{tab:mcmc_cases}, along with refitted cosmological parameters. The fiducial scenario A0 represents zero coupling ($\beta=0$) and a scalar field potential slope of $\lambda=0.6$, which corresponds to the lower bound of the $1\sigma$ region allowed by the MCMC constraints (see Table \ref{tab:cosmo_params}). 

For each of the 10 cosmological configurations, we perform one N-body simulation. Each simulation consists of $1024^3$ cold dark matter (CDM) particles within a cubic box of size $L_{\text{box}} = 1000 \, h^{-1} \, \text{Mpc}$, where $h$ is the dimensionless Hubble constant defined by $H_0 = h \times 100 \,\text{km}\,\text{s}^{-1}\,\text{Mpc}^{-1}$. Snapshots are generated at $z = 0$.

DM halos are identified from these snapshots using the halo finder \texttt{ROCKSTAR} \cite{Behroozi2012}. The halo center is defined as the average position of particles in the innermost subgroup, and the halo velocity is the average velocity of particles within the innermost 10\% of the halo’s virial radius. The resulting halo catalogs, together with the original snapshots, are then used as tracers to construct void catalogs at $z = 0$, as described in Section~\ref{sec:velocity}.

\begin{table*}
\large
\centering
\begin{tabular}{c c c c c c c c c c}
\hline
\hline
No. & $\beta$ & $\lambda$ & $H_0\ [\mathrm{km\ s^{-1}\ Mpc^{-1}}]$ & $\Omega_b$ & $\Omega_c$  & $n_s$ & $A_s\ [10^{-9}]$ & $\gamma_1$ & $\gamma_2$ \\
\hline
A0 & 0 & 0.6 & 67.21 & 0.0494 & 0.2602  & 0.9679 & 2.115 & 1 & 1 \\
\hline
A1 & $-$1.4 & 0.6 & 67.78 & 0.0485 & 0.2568  & 0.9670 & 2.111 & 1.4281 & 0.9021 \\
A2 & $-$1.4 & 1 & 67.42 & 0.0490 & 0.2588 & 0.9676 & 2.116 & 1.9779 &  0.7711 \\
A3 & $-$1.4 & 1.4 & 66.86 & 0.0499 & 0.2623  & 0.9685 & 2.122 & 2.4931 & 0.6379 \\
\hline
A4 & $-$0.8 & 0.6 & 67.68 & 0.0487 & 0.2574 & 0.9671 & 2.112 & 1.3597 &  0.9173 \\
A5 & $-$0.8 & 1 & 67.15 & 0.0495 & 0.2607 & 0.9681 & 2.117 & 1.8266 & 0.8032 \\
A6 & $-$0.8 & 1.4 & 66.34 & 0.0508 & 0.2659 & 0.9693 & 2.125 & 2.2611 & 0.6834 \\
\hline
A7 & $-$0.2 & 0.6 & 67.44 & 0.0490 & 0.2588 & 0.9676 & 2.114 & 1.1703 & 0.9602 \\
A8 & $-$0.2 & 1 & 66.43 & 0.0507 & 0.2657 & 0.9691 & 2.123 & 1.3977 & 0.9007 \\
A9 & $-$0.2 & 1.4 & 64.86 & 0.0532 & 0.2767  & 0.9710 & 2.137 & 1.6028 & 0.8330\\
\hline
\hline
\end{tabular}
\caption{Refitted parameters employed in the simulations. $\beta$ is the momentum coupling constant, and $\lambda$ is the scalar field potential slope. $H_0$ is the Hubble constant, and $\Omega_{x}$ are the energy density parameters, with $x\in\{b,c\}$, representing baryons and CDM, respectively. The spectral index $n_s$ and the amplitude $A_s$ quantify the properties of primordial scalar perturbations. Lastly, $\gamma_1$ and $\gamma_2$ are the friction and gravitational coefficients at $z=0$, respectively.}
\label{tab:mcmc_cases}
\end{table*}

\section{Methodology} \label{sec:velocity}
    \subsection{VOID finder}

To identify voids using halo and CDM tracers, we employed the Void Identification and Examination (\texttt{VIDE}) toolkit \cite{Sutter2015}, a modified version of \texttt{ZOBOV} \cite{Neyrinck2008}. \texttt{VIDE} performs a Voronoi tessellation of the tracers to locate minima in the density field. Around these minima, it applies a watershed transform \cite{Platen2007} to identify adjacent tessellation cells containing tracers with monotonically increasing densities, finding basins of underdensity to uncover a hierarchical tree of subvoids and voids.

In our calculations, we define voids as regions with a density contrast $\delta_c \leq -0.8$. This threshold is based on the spherical expansion model for an underdense region in an Einstein-de Sitter Universe, where $\delta_c \leq -0.8$ denotes the density contrast inside the top hat at the point of shell crossing, when matter accumulates at the void’s edge to form a denser wall \cite{Blumenthal1992, Sheth2004, Neyrinck2008}.

The center of a void, $\mathbf{X}_{\mathrm{v}}$, is defined as the volume-weighted average position of its tracers, where the weight for each tracer is given by its Voronoi cell volume $V_j$ \cite{Sutter2015}:
\begin{equation} \textbf{X}_{\mathrm{v}} = \frac{\sum_j \textbf{x}_j V_j}{\sum_j V_j} .  \end{equation}

Here, $\textbf{x}_j$ denotes the position of the $j$-th tracer. Since the voids identified by \texttt{VIDE} are generally non-spherical, the Voronoi cell with the lowest density is not necessarily located at the void’s center \cite{Schuster2023}. Therefore, we define an effective radius for each void as

\begin{equation}
r_{\text{v}} = \left( \frac{3}{4\pi} \sum_{j} V_{j} \right)^{1/3},
\end{equation}
where the void volume is given by the sum of the non-overlapping Voronoi cell volumes $V_{j}$ of all tracers that constitute the void. Furthermore, \texttt{VIDE} includes a density-based merging threshold for voids, which is a free parameter. This threshold specifies the highest density (in units of the mean number density of tracers, $\bar{n}$) allowed along the ridge line separating a void and an adjacent basin for them to merge \cite{Sutter2015, Blumenthal1992, Schuster2023}. When merging occurs, one void becomes the parent and the other a sub-void, creating a hierarchical structure. Following \cite{Sutter2015, Schuster2023}, we adopt a very low threshold of $10^{-9}\,\bar{n}$, effectively disabling merging. This ensures that only basins connected by extremely low-density paths could merge, preserving the physical integrity of void boundaries and avoiding unrealistic bridging through dense regions.

\subsection{Catalogs}

In our study, we use halos and CDM particles as tracers to identify voids at $z=0$. The voids identified using halo and CDM  tracers are referred to as halo voids and CDM voids, respectively. For each category, the corresponding tracer is used to compute the velocity profiles. The mean number densities of halo and CDM tracers for the 10 scenarios, given in Table \ref{tab:void_stats}, are of the same order of magnitude as the mean tracer density in state-of-the-art galaxy surveys, such as those expected from \textit{Euclid} \cite{Hamaus2022}. 

Based on our simulation setup, to ensure consistency between the  halo and CDM void catalogs, we sample approximately 0.41\% of the CDM particles, matching the number of CDM tracers to that of the halo tracers. This procedure eliminates differences in statistics arising from different mean tracer number densities: the halo and CDM voids would be of similar sizes, and the mean tracer separation $\bar{r}_{t}$ is similar in the two tracer sets.

\begin{table*}
\centering

\resizebox{\textwidth}{!}{%
\begin{tabular}{|c|c|c|c|c|c|c|c|c|}
\hline
$M_{\text{cut}}$ [$10^{12} M_{\odot}/h$] & $N_h$ [$10^6$] & $\bar{n}_h$ [$10^{-3} (\text{Mpc}/h)^{-3}$] & $N_{\text{c}}$ [$10^6$] & $\bar{n}_{\text{c}}$ [$10^{-3} (\text{Mpc}/h)^{-3}$] & $\bar{r}_t (\text{halos})$ [Mpc/$h$] & $\bar{r}_t (\text{CDM})$ [Mpc/$h$] & $N_\mathrm{v}$ (halos) [$10^3$] & $N_\mathrm{v}$ (CDM) [$10^3$] \\
\hline
$1.0$ & $4.3$--$4.4$ & $4.3$--$4.4$ & $4.4$ & $4.4$ & $6.1$--$6.2$ & $6.1$ & $18$--$19$ & $23$--$24$ \\
\hline
\end{tabular}%
}

\caption{Void catalog statistics at $z = 0$ across 10 simulation scenarios, with the halo mass threshold  $M_{\text{cut}}$, number of tracers $N_{h/c}$, mean tracer number density  ${\bar n_{h/c}}$, mean tracer separation  ${\bar r}_t$, and number of voids $N_\text{v}$ for both the halo ($h$) and CDM ($\text{c}$) distributions.}
\label{tab:void_stats}
\end{table*}

\subsection{Radial velocity profile }

To quantify the radial motion of tracers around a void, we compute the volume-weighted average radial velocity profile. Because voids vary in size, we compute this profile for the $i$-th void as a function of the scaled distance $x = r/r_\text{v}^{(i)}$ \cite{Hamaus2014, Schuster2023}: 

\begin{equation}
    v_v^{(i)}(x) = \frac{\sum_j \mathbf{v}_j \cdot \mathbf{\hat{r}}_j V_j \Theta(r_j)}{\sum_j V_j \Theta(r_j)},
    \label{eq:v_indiv} 
\end{equation}
where $\bm{v}_j$ is the peculiar velocity of tracer $j$, $\bm{\hat{r}}_j$ $(r_j)$ is the unit vector (distance) from the void center to tracer $j$ and the function  $\Theta(r_j) = \vartheta\big[ \frac{r_j}{r_\text{v}^{(i)}} - (x - \delta x) \big] \vartheta\big[ -\frac{r_j}{r_\text{v}^{(i)}} + (x + \delta x) \big]$ is the product of two Heaviside step functions $\vartheta$ that selects tracers within the scaled radial bin  $[x - \delta x, x + \delta x]$, where $\delta x$ represents the half-width of the scaled bin. The denominator represents the total Voronoi volume of all tracers enclosed within this specific scaled radial bin. We then stack $v^{(i)}_\text{v}(x)$ by averaging the profiles of individual voids within a catalog. This stacked velocity profile is computed using the \textit{individual stack} method, defined as \cite{Schuster2023}

\begin{equation}
    v_\text{v}(x) = \frac{1}{N_{\text{v,bin}}} \sum_{i \in \text{bin}} v^{(i)}_\text{v}(x), 
    \label{eq: Individual_v_rad}
\end{equation}
 where $N_\text{v,bin}$ is the number of voids in that specific radius bin. The statistical uncertainties on these stacked profiles are estimated using Jackknife resampling over the void catalogs to account for the covariance between overlapping void volumes at large distance from the void center \cite{Schuster2023}. To capture the dynamics of tracers both inside and outside a void, the profile is computed out to three times the effective radius of each void. Furthermore, to prevent the smearing of features that occurs when averaging voids of vastly different sizes,  we perform a two-step stacking procedure. First, we stack within narrow bins of $r_\text{v}$. This is  necessary because void dynamics inherently depend on their absolute size; small voids tend to be overcompensated and are prone to collapse, whereas large voids are undercompensated and expand continuously. Stacking across all sizes would artificially smear these distinct physical signatures. Second, within these narrow bins, the individual profiles are evaluated and averaged in terms of the scaled distance $x=r/r_{\text{v}}$.  Because voids still vary in size within a given bin, scaling ensures that geometric features, such as the compensation wall, are consistently located at $x\sim1$, preventing the structural smearing that would occur if they were averaged using physical distance $r$. Lastly, to minimize resolution effects and ensure the accuracy of the profiles, we restrict our analysis to voids with an effective radius greater than the mean tracer separation, $r_\text{v} > \bar{r}_{t}$.


By definition, a positive velocity indicates an outflow of tracers from the void center, which typically occurs within the effective radius of the void, while a negative velocity indicates an inflow toward the void center.

\subsection{Radial velocity dispersion profile }

The radial velocity dispersion around a void quantifies the spread of tracer radial velocities relative to the void’s mean radial flow. For the $i-$th void, the radial velocity dispersion is defined as the volume-weighted standard deviation of the radial velocity components of all tracers located at a scaled distance $x$ from the void center \cite{Hamaus2015}:


\begin{equation}
\sigma_\text{v}^{(i)}(x) =
\sqrt{
\frac{\sum_{j} \Bigr[\bm{v}_j\cdot \bm{\hat{r}}_j - v^{(i)}_\text{v}(x)\Bigr]^2  \, V_j \Theta (r_j)}{\sum_j V_j \Theta(r_j)} 
}.
\label{eq:per_void_dispersion}
\end{equation}

Similar to Eq.~\eqref{eq: Individual_v_rad}, the stacked velocity dispersion profile is defined as 

\begin{equation}
    \sigma_\text{v}(x) = \frac{1}{N_\text{v,bin}} \sum_{i \in \text{bin}} \sigma^{(i)}_\text{v}(x).
\end{equation}

Although less studied than $v_{\text{v}}(r)$, $\sigma_{\text{v}}(r)$ is a key observable in the formation of LSS. It contributes to the modeling of RSD in void–galaxy correlations, where the dispersion acts as a small-scale smearing term in anisotropic clustering analyses \cite{Cai2016}.  Moreover, it is sensitive to fifth-force effects in modified gravity: simulations show enhanced dispersion relative to general relativity in unscreened regions, making it a useful probe for theories such as  $f(R)$ gravity \cite{Cai2015, Wilson2023}.

Physically, $\sigma_{\text{v}}(r)$ measures the random motions of tracers and follows a characteristic shape in void environments. Near the void center, the density is very low, so there are few tracers and minimal gravitational interactions, resulting in small random motions and thus low dispersion. As $r$ increases toward the void edge $r_\text{v}$, the tracer density rises, and stronger gravitational interactions lead to greater random motions, causing $\sigma_{\text{v}}(r)$ to increase and peak near the void boundary. Beyond the void edge, the environment approaches the mean cosmic density, so $\sigma_{\text{v}}(r)$ stabilizes at a value determined by the typical random motions of tracers in the background.

In \cite{Luo2026}, the parameters $\beta$ and $\lambda$ of the Type 3 model were shown to strongly influence matter pairwise velocity and velocity dispersion, highlighting their potential as probes of DE models. Motivated by these findings and the fact that cosmic voids serve as an ideal environment for detecting DE due to the minimal influence of complex baryonic physics, we extend the investigation to examine the dependence of void radial velocity profile statistics on $\beta$ and $\lambda$.

\section{Results} \label{sec:results}

This section presents the results for $v_\text{v}(x)$ and $\sigma_\text{v}(x)$ at redshift $z = 0$, obtained from the simulation sets with a box size of $L_{\text{box}} = 1000 \, h^{-1} \, \text{Mpc}$, as described in Section~\ref{sec:velocity}. The scalar field and its momentum coupling to DM introduce three key effects relative to $\Lambda$CDM: (i) a dynamical modification to the expansion history of the Universe driven by the evolving scalar field, (ii) an additional cosmological friction force and a reduced gravitational force acting on DM, and (iii) refitted cosmological parameters required to match observational data. According to \cite{chamings2020}, a faster expansion of the universe mildly suppresses structure growth by reducing the magnitude of the metric perturbation term $|\dot{\Phi}|$ in Eq.~\eqref{perturbations}. Conversely, the altered cosmological friction and the refitted cosmological parameters are more significant in modifying the structure formation.
\subsection{Void radial velocity profile analysis}

\label{sec: void radial velocity profile analysis}

\begin{figure*}
    \includegraphics[width=.80\textwidth]{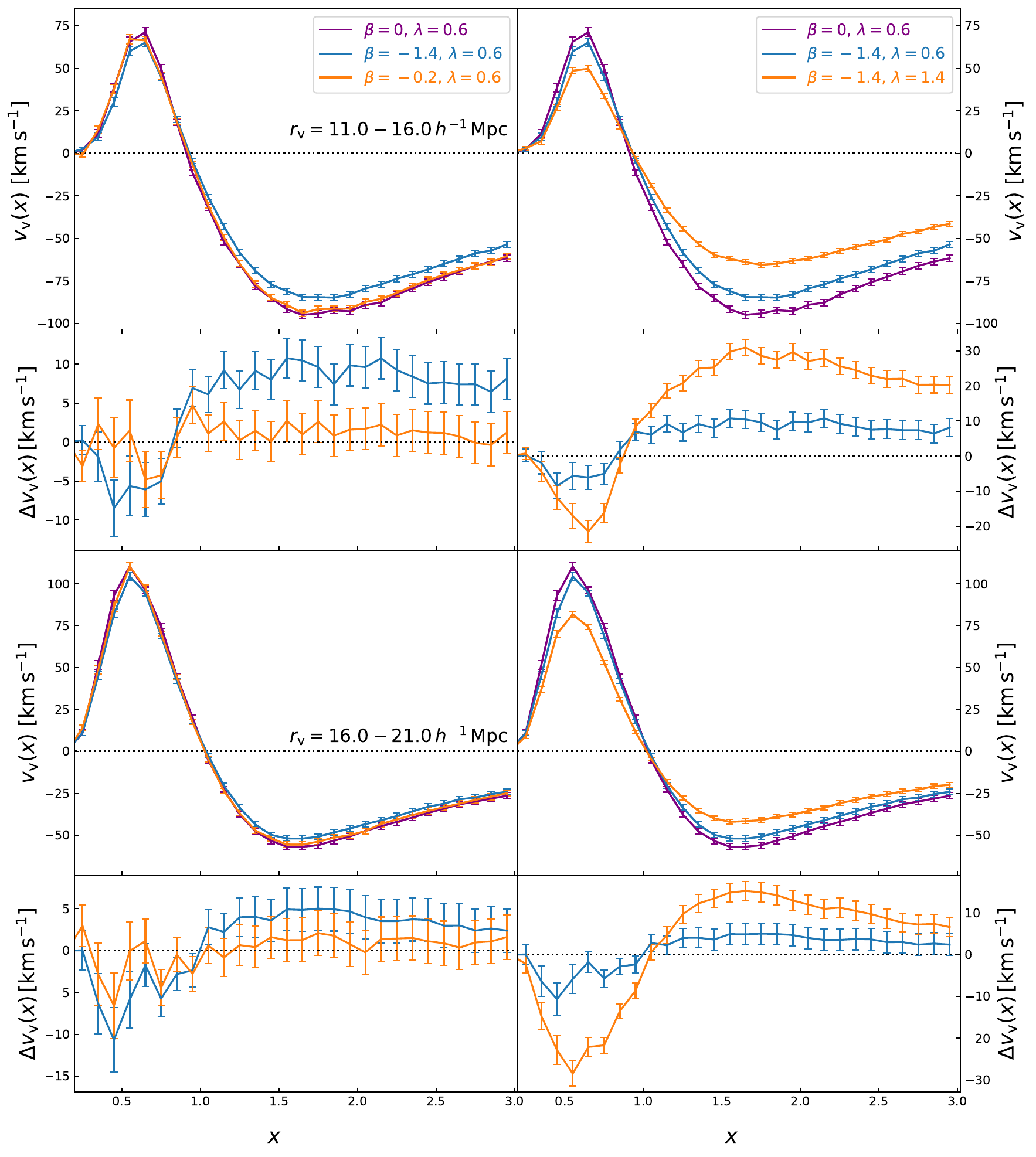}
    \caption{ $v_\mathrm{v}$ of isolated halo voids of isolated halo voids as a function of the scaled distance $x=r/r_\text{v}$ for $r_\text{v} = 11 $--$ 16\,h^{-1} \,\text{Mpc}$ (top row) and $r_\text{v} = 16 $--$ 21\,h^{-1}\,\text{Mpc}$ (third row).  The left and right columns show the velocity  profiles for varying $\beta$ (with $\lambda=0.6$) and $\lambda$ (with $\beta=-1.4$), respectively, compared to the fiducial case A0 with $\beta = 0, \, \lambda = 0.6$ (purple lines). Each $r_{\text{v}}$ bin contains 5000 to 5500 voids. The second row from top and bottom row display the difference (in $\mathrm{km \,s^{-1}}$) of $v_\text{v}$ from that of A0, $\Delta v_{\mathrm{v}}= v_{\mathrm{v}} (\text{other cases}) - v_{\mathrm{v}} (\mathrm{A0})$. }\label{fig:vel_h}
\end{figure*}

Figures \ref{fig:vel_h} and \ref{fig:vel_p} show $v_\text{v}(x)$ in the top subpanels for halo and CDM voids, respectively, and their differences (in $\mathrm{km \,s^{-1}}$) from the fiducial scenario A0 with no momentum exchange ($\beta = 0$) in the bottom subpanels. Each figure is organized into top and bottom rows, corresponding to $r_\text{v} = 11 $--$ 16\, h^{-1} \, \text{Mpc} $ and $r_\text{v} = 16 $--$ 21\,  h^{-1} \, \text{Mpc}$, respectively. Furthermore, the left (right) column illustrates the impact of varying $\beta \, (\lambda)$ at a fixed $\lambda = 0.6 \,(\beta = -1.4)$.

From the figures, we observe a characteristic behavior in $v_{\text{v}}(x)$ around void centers. At small separations ($x \ll 1$), tracers exhibit coherent outflows driven by the expansion of the underdense interior of the void \cite{Cai2015}. $v_{\text{v}}(x)$ steadily rises from zero at the center and reaches its maximum inside $r_\text{v}$, where the influence of the surrounding overdense shell—the compensation wall—begins to dominate. This wall forms as matter is evacuated from the void interior and accumulates near its boundary, partially compensating for the central underdensity. When matter passes through the wall, the gravitational pull of the wall continuously reduces the outflow, decreasing the net tracer velocity to zero near $ x \sim 1 $. Beyond this scale, tracer motion transitions to inflow toward the void, with the inflow velocity peaking around $x \sim 1.5$. At larger separations, the wall’s influence weakens, and $ v_{\text{v}}(x) \rightarrow 0 $, conforming to the Hubble flow \cite{Hamaus2014, Schuster2023}.

$v_{\text{v}}(x)$ shifts up with the void size.  Smaller voids experience stronger effects from the compensation wall because the wall’s mass typically exceeds the mass deficit evacuated from the void's interior \cite{Sheth2004}. This leads to a steeper transition in $v_{\text{v}}(x)$, as the wall’s gravitational pull more readily overcomes the void's interior expansion, facilitating tracer inflow beyond the void radius \cite{Cai2014, Paz2013, Ceccarelli2013} and gradually leading to void collapse. In contrast, larger voids tend to be undercompensated, with the wall’s mass insufficient to fully offset the interior mass deficit. Their surrounding walls are insufficient to reverse the outflow. Consequently, their $v_{\text{v}}(x)$ profiles feature stronger outflows that peak near the compensation wall and only approach zero at large distances from the void.

From the top panels of Figure~\ref{fig:vel_h} and \ref{fig:vel_p}, for both the halo and CDM voids, all cases show a lower magnitude of $v_\text{v}(x)$ compared to the uncoupled case  (purple line) with $\lambda=0.6$, corresponding to friction and gravitational coefficients, $\gamma_1 = 1$ and $\gamma_2 = 1$, respectively. This trend mirrors that observed in the mean matter pairwise peculiar velocity for the Type 3 model in the linear regime  \cite{Luo2026}, where a net transfer of momentum from DM particles to the scalar field ($\beta$ < 0) introduces additional friction on the tracers, as indicated by the increased friction coefficient ($\gamma_1$ > 1) relative to the uncoupled case (see Table \ref{tab:mcmc_cases}). As a result, the tracer motion is damped, suppressing outflow inside the void and inflow beyond the void radius.

Furthermore, weaker coupling (smaller $|\beta|$) reduces the frictional damping of radial motions, leading to larger $|v_\text{v}(x)|$ relative to strongly coupled cases. Conversely, a steeper potential (larger $\lambda$) enhances friction, slowing tracer motion and further reducing  $|v_\text{v}(x)|$.

To better understand how $\beta$ and $\lambda$ influence tracer dynamics around voids, we compute the average radial velocity span,  ${\bar v}_{\text{span}} =  \max(v_\text{v}(x)) - \min(v_\text{v}(x))$, where the maximum (minimum) velocity is calculated as the average of the three maximum (minimum) velocity values in each profile. This velocity span is computed for both halo and CDM voids, denoted as $ \bar{v}_{\mathrm{span}}^{\mathrm{h}}$ and $\bar{v}_{\mathrm{span}}^{\mathrm{p}}$, respectively. As $\beta$ increases towards 0, tracers lose less momentum to the scalar field, increasing $ \bar{v}_{\mathrm{span}}$. Conversely, increasing $\lambda$ dampens tracer motion, thereby decreasing $ \bar{v}_{\mathrm{span}}$.

To quantify the impact of the dark-sector interaction, we evaluate the fractional deviation of $ \bar{v}_{\mathrm{span}}$ relative to the fiducial scenario (A0). We define this deviation as:
\begin{equation}
    \Delta  \bar{v}^{\text{[h/p]}}_{\text{span}}(\beta, \lambda) = \frac{\bar{v}^{\text{[h/p]}}_{\text{span}}(\beta, \lambda)}{\bar{v}^{\text{[h/p]}}_{\text{span}}(0, 0.6)} - 1 .
    \label{eq:v_span_dev}
\end{equation}

To provide a robust, observationally useful diagnostic, we fit the fractional deviations across our simulation grid using a quadratic Taylor expansion. Centering the expansion on the fidicual scenario A0 ($\beta = 0$, $\lambda = 0.6$), the regression model is given by:
\begin{equation}
    \Delta \bar{v}^{[\text{h/p}]}_{\text{span}} = c_\beta \Delta\beta + c_\lambda \Delta\lambda + c_{\beta\beta} \Delta\beta^2 + c_{\beta\lambda}\Delta\beta \Delta\lambda,
    \label{eq:quad_fit}
\end{equation}
where $\Delta\beta = \beta$ and $\Delta\lambda = \lambda - 0.6$. While a complete quadratic expansion would include a $\Delta\lambda^2$ term, empirical testing reveals its coefficient to be nearly zero for our parameter space, so this term is omitted.

Furthermore, to quantify the goodness of fit, we evaluate the coefficient of determination, $R^2$, which measures the proportion of the variance in $ \Delta  \bar{v}_{\text{span}}$ explained by the dark-sector parameters. Given our sample size of 9 simulation scenarios fitted with $4$ parameters, we also compute the adjusted $R^2$ ($\bar{R}^2$). This metric incorporates a strict penalty for over-parameterization, allowing us to investigate and rule out potential overfitting of the regression model.

The best-fit coefficients for the halo and CDM voids are summarized in Table \ref{tab:coefficients_h} and Table \ref{coefficients_c}, respectively. The high $\bar{R}^2$ values (close to 1) demonstrate that this quadratic model accurately captures the parameter dependencies.

For both types of voids of different sizes, the linear coefficients ($c_\beta > 0$ and $c_\lambda < 0$) in the tables confirm our qualitative findings: weaker momentum coupling and a steeper potential induce opposite primary effects on $ \Delta \bar{v}_{\text{span}}$. However, the significant non-zero cross-terms ($c_{\beta\beta}$) indicate that the parameters are entangled. Based on Eq.~\eqref{scf_lag}, this cross-dependence arises because the momentum interaction between the dark components scales with $\dot{\phi}^2$. A steeper potential drives faster evolution of the scalar field $\dot{\phi}$, rendering $v_\text{v}(x)$ more sensitive to the momentum coupling.

Additionally, within the parameter space explored in our simulations, the magnitude of $ \Delta \bar{v}_{\text{span}}$ reaches $\sim 30\%$ for scenario A3 ($\beta=-1.4$, $\lambda=1.4$), which corresponds to the cosmology with the strongest momentum coupling and steepest scalar potential considered in this work. This substantial deviation demonstrates the potential of using void radial velocity statistics to probe the Type 3 model in upcoming observational surveys.

\begin{table*}[t]

\centering
\renewcommand{\arraystretch}{1.3}
\begin{tabular}{l l c c c c c}
\hline\hline
Quantity & ${r_\text{v}}$ [$h^{-1}$\text{Mpc}] & $c_\beta \, (\%) $ & $c_\lambda \, (\%) $ & $c_{\beta\beta} \, (\%) $ & $c_{\beta\lambda} \, (\%) $ & $R^2 (\bar{R}^2)$ \\
\hline
\multirow{2}{*}{$\Delta \bar{v}_{\text{span}}$} 
& 11.0 -- 16.0 & $14.33 \pm 2.86$ & $-12.79 \pm 2.20$ & $6.16 \pm 2.20$ & $10.54 \pm 2.44$ & 0.995 (0.992) \\
& 16.0 -- 21.0 & $11.81 \pm 2.63$ & $-8.61 \pm 2.03$ & $6.07 \pm 2.02$ & $13.64 \pm 2.24$ & 0.971 (0.953)\\
\cline{1-7}
\multirow{2}{*}{$\Delta \bar{\sigma}_{\text{span}}$} 
& 11.0 -- 16.0 & $13.65 \pm 1.05$ & $-13.78 \pm 0.81$ & $5.98 \pm 0.81$ & $9.78 \pm 0.89$ & 0.998 (0.997)\\
& 16.0 -- 21.0 & $12.25 \pm 1.32$ & $-11.43 \pm 1.01$ & $5.93 \pm 1.02$ & $11.92 \pm 1.12$ & 0.988 (0.981)\\
\hline\hline

\end{tabular}

\caption{Best-fit coefficients for the regression of $\Delta \bar{v}_{\text{span}}$ and $\Delta \bar{\sigma}_{\text{span}}$ for halo voids relative to the uncoupled A0 scenario. The final column provides the $R^2$ and $\bar{R}^2$ for each fit.}
\label{tab:coefficients_h}
\end{table*}

\subsection{Void radial velocity dispersion profile analysis}

\begin{figure*}
    \includegraphics[width=.80\textwidth]{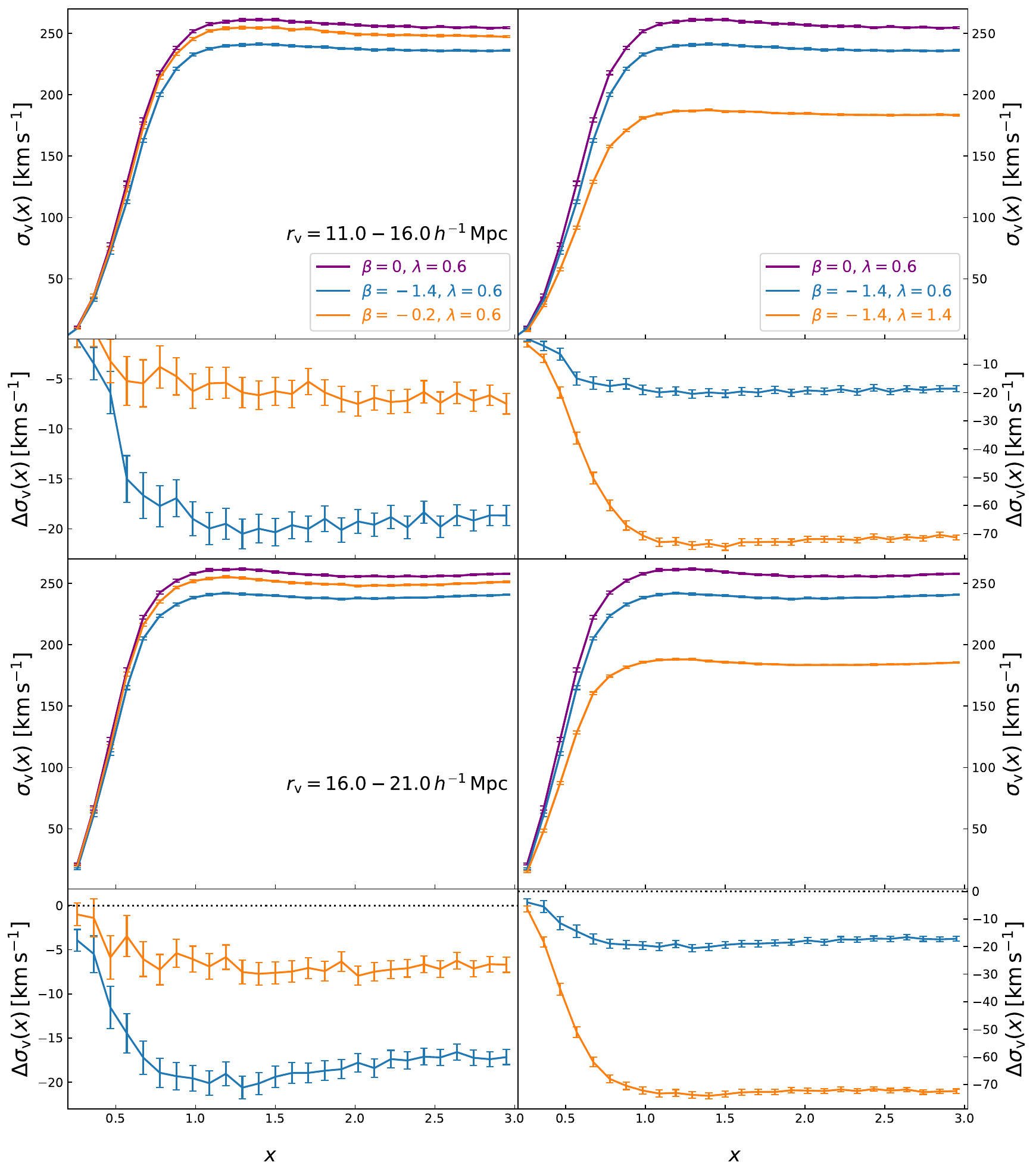}
    \caption{$\sigma_\mathrm{v}$ of isolated halo voids as a function of the scaled distance $x=r/r_\text{v}$ for $r_\text{v} = 11 $--$ 16\,h^{-1} \,\text{Mpc}$ (top row) and $r_\text{v} = 16 $--$ 21\,h^{-1}\,\text{Mpc}$ (third row from top).  The left and right columns show the profiles for varying $\beta$ (with $\lambda=1.4$) and $\lambda$ (with $\beta=-1.4$), respectively, compared to the fiducial case A0 with $\beta = 0, \, \lambda = 0.6$ (purple lines). Each $r_\text{v}$ bin contains 5000 to 5500 voids. The second row from the top and the bottom row display the difference  (in $\mathrm{km \,s^{-1}}$) from the fiducial configuration's (A0), $\Delta \sigma_{\mathrm{v}}= \sigma_{\mathrm{v}} (\text{other cases}) - \sigma_{\mathrm{v}} (\mathrm{A0})$.}\label{fig:sigma_h}
\end{figure*}

Similar to Figure \ref{fig:vel_h} and \ref{fig:vel_p}, Figure \ref{fig:sigma_h} and \ref{fig:sigma_p} present $\sigma_\mathrm{v}(x)$ for the halo and CDM voids, respectively. At small separations from the void center, the random motions of tracers are minimal due to the shallow and smooth gravitational potential in the underdense core. However, because the matter density rises near the compensation wall, stronger gravitational interactions cause random motions to grow and eventually stabilize at a constant value at large separations.

Comparing $\sigma_{\text{v}}(x)$ with the matter pairwise velocity dispersion provides insight into the contrasting dynamics of underdense and overdense environments. In contrast to the former, the latter peaks at small inter-tracer separations, coinciding with regions of strong gravity and virialized motions within DM halos, and then gradually decreases to stabilize to a constant value at large separations as the dynamics transitions from the local non-linear regime to the linear large-scale flow \cite{zhang2024, Luo2026}.


It was shown in \cite{Hamaus2015} that larger voids can exhibit a higher \(\sigma_\text{v}(x)\). This is purely an artifact of the rescaling by the void effective radius $r_{\text{v}}$: because $\sigma_\text{v}$ itself increases with physical distance $r$ from the void center, the same scaled coordinate $ r/r_v$ corresponds to a larger $r$ for bigger voids. In fact, at fixed physical distance, the dispersions are very similar across different void sizes. Crucially, because our analysis compares the Type 3 model against the uncoupled scenario within the same $r_\text{v}$ bins, the coordinate scaling is identical across all scenarios. Therefore, the discrepancies between different scenarios in $\sigma_\text{v}(x)$ are physical signatures of the dark-sector interactions, safely isolated from any scaling effects.

Similar to $v_\text{v}(x)$, Figures~\ref{fig:sigma_h} and \ref{fig:sigma_p} show that all cases exhibit suppressed $\sigma_\text{v}(x)$ relative to the uncoupled case A0. This suppression arises from the net momentum transfer from DM particles to the scalar field, which introduces additional friction on the tracers. Furthermore, there is a clear dependence of $\sigma_\text{v}(x)$ on $\beta$ and $\lambda$. From the left panels of Figures~\ref{fig:sigma_h} and \ref{fig:sigma_p}, weaker momentum coupling (smaller $|\beta|$) reduces the frictional damping, leading to stronger fluctuations across all separation scales from the void center. In contrast, a steeper potential (larger $\lambda$) enhances friction, further damping tracer motions and reducing $\sigma_\text{v}(x)$. 

This trend becomes more evident when examining the average radial velocity dispersion span, ${\bar \sigma_{\text{span}}} =  \max(\sigma_\text{v}(x)) - \min(\sigma_\text{v}(x))$, the difference between the weighted average maximum and minimum $\sigma_\mathrm{v}(x)$.  Following the previous analysis of $v_\text{v}(x)$, we quantify the effects of the Type 3 model on  $\sigma_{\text{v}}(x)$ by evaluating the fractional deviation of ${\bar \sigma_{\text{span}}}$:
\begin{equation}
    \Delta \bar{\sigma}^{[\text{h/p}]}_{\text{span}}(\beta, \lambda) =  \frac{\bar{\sigma}^{[\text{h/p}]}_{\text{span}}(\beta, \lambda)}{\bar{\sigma}^{[\text{h/p}]}_{\text{span}}(0, 0.6)} - 1.
    \label{eq:sigma_span_dev}
\end{equation}
Similar to  $\Delta \bar{v}_{\text{span}}$, we fit these deviations using the 4-parameter quadratic regression model established in Eq.~(\ref{eq:quad_fit}). The resulting coefficients for $\Delta \bar{\sigma}_{\text{span}}$ for the halo and CDM voids are also provided in Table \ref{tab:coefficients_h} and Table \ref{coefficients_c}, respectively. 

From the tables, it is clear that because weaker momentum coupling increases the amplitude of fluctuations, ${\bar \sigma_{\text{span}}}$ increases $(c_\beta>0)$. Conversely, a steeper potential dampens tracer motion, reducing fluctuations and  ${\bar \sigma_{\text{span}}}$ $(c_\lambda<0)$.

Similar to the results for $\Delta \bar{v}_{\text{span}}$, the magnitude of $\Delta \bar{\sigma}_{\text{span}}$ reaches $\sim 30\%$ for scenario A3.

\hspace{3cm}

\section{Conclusions} \label{sec:conclusions}

In this paper, we present a quantitative analysis of the effects of the momentum coupling $\beta \, (< 0)$ and the potential slope $\lambda$ in the Type 3 model on the radial velocity $v_\text{v}^{\text{[p/h]}}$ and velocity dispersion profile $\sigma_\text{v}^{\text{[p/h]}}$ around voids. These effects arise from a combination of the altered cosmological friction force acting on DM, the refitted cosmological parameters, and a dynamical modification to the cosmological expansion history driven by the evolving scalar field.

Consistent with the results in \cite{Luo2026}, the degeneracy between $\beta \, (< 0)$ and $\lambda$ observed in the mean matter pairwise peculiar velocity statistics is also evident in the radial velocity statistics of voids across different regimes (inside and around the void) and for various void sizes. Our key conclusions are summarized below:

\hspace*{\fill}

(i) For both CDM and halo voids, increasing $\beta$ (decreasing  $|\beta|$, weaker coupling) and $\lambda$ (steeper potential) induce opposite primary effects on the void radial velocity profile statistics. However, as evidenced by the  cross-terms in our quadratic regression for $\Delta \bar{v}_{\text{span}}$ and $\Delta \bar{\sigma}_{\text{span}}$, these parameters do not act independently; a steeper potential drives faster evolution of the scalar field, which dynamically amplifies the sensitivity of the velocity statistics to the momentum coupling.

(ii) The effects induced by these parameters are highly consistent across voids of different sizes. Quantitatively, the Type 3 model parameters induce systematic fractional deviations in $\Delta \bar{v}_{\text{span}}$ and $\Delta \bar{\sigma}_{\text{span}}$, reaching up to $\sim30\%$
within the parameter space explored relative to the fiducial scenario. This fractional deviation is comparable to our previous findings for matter pairwise velocity statistics under the same fiducial scenario \cite{Luo2026}. This similar sensitivity to IDE parameters establishes the void radial velocity statistics as a complementary probe of IDE. We provide a 4-parameter quadratic regression model that accurately captures this non-linear parameter dependence, highlighting the potential of void velocity statistics as a sensitive, observationally accessible probe of IDE.


Although we visually depict the underdense region in the proximity of the void centers ($x = [0, 0.5]$) in our plots, we do not perform a quantitative analysis in this range due to limited resolution and tracer density in our simulations. Future studies would benefit from higher-resolution simulations of the Type 3 model to better characterize void dynamics near the void core.
\section*{Acknowledgements} \label{sec:acknowledgements}

We would like to thank Nico Hamaus and Nico Schuster for the helpful advices and feedbacks on utilising \texttt{VIDE} \cite{Sutter2015}. This research is partially funded through grants awarded by the Research Grants Council of the Hong Kong Special Administrative Region, China (Project Numbers 14301214 and AoE/P-404/18). We also extend our gratitude to the CUHK Central High Performance Computing Cluster for providing the computational resources used in this study. K.C.C. acknowledges the support by the National Science Foundation of China under grant Nos. 12273121 and 12533002, and the science research grant from the China Manned Space Project with CMS-CSST-2025-A02. 

\bibliographystyle{apsrev4-2}
\nocite{*}

\bibliography{references}

@article{Chan_etal2019,
       author = {{Chan}, Kwan Chuen and {Hamaus}, Nico and {Biagetti}, Matteo},
        title = "{Constraint of void bias on primordial non-Gaussianity}",
      journal = {"Phys. Rev. D"},
         year = 2019,
        month = jun,
       volume = {99},
       number = {12},
          eid = {121304},
        pages = {121304},
          doi = {10.1103/PhysRevD.99.121304},
archivePrefix = {arXiv},
       eprint = {1812.04024},
 primaryClass = {astro-ph.CO},
       adsurl = {https://ui.adsabs.harvard.edu/abs/2019PhRvD..99l1304C}
}

@article{di_valentino2017,
    author = "Di Valentino, Eleonora and Melchiorri, Alessandro and Mena, Olga",
    title = "{Can interacting dark energy solve the $H_0$ tension?}",
    doi = "10.1103/PhysRevD.96.043503",
    journal = "Phys. Rev. D",
    volume = "96",
    number = "4",
    pages = "043503",
    year = "2017",
    month = aug
}

@article{Blumenthal1992,
  author = {Blumenthal, G. R.  and others},
  title = {The Largest Possible Voids},
  journal = {Astrophys. J.},
  volume = {388},
  pages = {234},
  year = {1992},
  month = apr,
  doi = {10.1086/171147},
  url = {https://doi.org/10.1086/171147}
}

@article{Sheth2004,
  author = {Sheth, Ravi K. and van de Weygaert, Rien},
  title = {A hierarchy of voids: much ado about nothing},
  journal = {Mon. Not. Roy. Astron. Soc.},
  volume = {350},
  number = {2},
  pages = {517--538},
  year = {2004},
  month = may,
  doi = {10.1111/j.1365-2966.2004.07661.x},
  url = {https://doi.org/10.1111/j.1365-2966.2004.07661.x}
}

@article{Schuster2023,
  author = {Schuster, Nico and others},
  title = {Why cosmic voids matter: nonlinear structure \& linear dynamics},
  journal = {JCAP},
  volume = {2023},
  number = {05},
  pages = {031},
  year = {2023},
  month = may,
  doi = {10.1088/1475-7516/2023/05/031},
  url = {https://doi.org/10.1088/1475-7516/2023/05/031}
}

@article{Sutter2015,
  author = {Sutter, P. S. and others},
  title = {VIDE: The Void IDentification and Examination toolkit},
  journal = {Astron. Comput.},
  volume = {9},
  pages = {1--9},
  year = {2015},
  doi = {10.1016/j.ascom.2014.10.002}
}

@article{Neyrinck2008,
  author = {Neyrinck, M. C.},
  title = {ZOBOV: a parameter-free void-finding algorithm},
  journal = {Mon. Not. Roy. Astron. Soc.},
  volume = {386},
  number = {4},
  pages = {2101--2109},
  year = {2008},
  doi = {10.1111/j.1365-2966.2008.13180.x}
}

@article{Luo2026,
  author        = {Luo, Kin Ho and Chu, Ming-chung and Zhang, Wangzheng},
  title         = {Probing the Type 3 interacting dark-energy model using matter pairwise velocity},
  journal       = {Phys. Rev. D},
  volume        = {113},
  number        = {8},
  pages         = {083511},
  year          = {2026},
  doi           = {10.1103/PhysRevD.113.083511},
  eprint        = {2508.04312},
  archivePrefix = {arXiv},
  primaryClass  = {astro-ph.CO},
}

@article{Platen2007,
  author = {Platen, Erwin and van de Weygaert, Rien and Jones, Bernard J. T.},
  title = {A cosmic watershed: the WVF void detection technique},
  journal = {Mon. Not. Roy. Astron. Soc.},
  volume = {380},
  number = {2},
  pages = {551--570},
  year = {2007},
  month = aug,
  doi = {10.1111/j.1365-2966.2007.12125.x},
  url = {https://doi.org/10.1111/j.1365-2966.2007.12125.x}
}

@article{Hamaus2014,
  author = {Hamaus, Nico and Sutter, P. M. and Wandelt, Benjamin D.},
  title = {Universal Density Profile for Cosmic Voids},
  journal = {Phys. Rev. Lett.},
  volume = {112},
  pages = {251302},
  year = {2014},
  month = jun,
  doi = {10.1103/PhysRevLett.112.251302},
  url = {https://doi.org/10.1103/PhysRevLett.112.251302}
}

@article{Cai2016,
  author = {Cai, Yan-Chuan and others},
  title = {Redshift-space distortions around voids},
  journal = {Mon. Not. Roy. Astron. Soc.},
  volume = {462},
  number = {3},
  pages = {2465--2477},
  year = {2016},
  month = {11},
  day = {01},
  doi = {10.1093/mnras/stw1809},
}

@article{Cai2015,
  author = {Cai, Yan-Chuan and Padilla, Nelson and Li, Baojiu},
  title = {Testing gravity using cosmic voids},
  journal = {Mon. Not. Roy. Astron. Soc.},
  volume = {451},
  number = {1},
  pages = {1036--1055},
  year = {2015},
  month = {07},
  day = {21},
  doi = {10.1093/mnras/stv777},
}

@article{Wilson2023,
  author = {Wilson, Caitlin and Bean, Rachel},
  title = {Challenges in constraining gravity with cosmic voids},
  journal = {Phys. Rev. D},
  volume = {107},
  number = {12},
  pages = {124008},
  year = {2023},
  doi = {10.1103/PhysRevD.107.124008}
}

@article{Cai2014,
  author = {Cai, Yan-Chuan and others},
  title = {A Possible Cold Imprint of Voids on the Microwave Background Radiation},
  journal = {Astrophys. J.},
  volume = {786},
  number = {2},
  pages = {110},
  year = {2014},
  month = apr,
  publisher = {The American Astronomical Society},
  doi = {10.1088/0004-637X/786/2/110},
  bibcode = {2014ApJ...786..110C}
}

@article{Paz2013,
  author = {Paz, Dante and others},
  title = {Clues on void evolution--II. Measuring density and velocity profiles on SDSS galaxy redshift space distortions},
  journal = {Mon. Not. Roy. Astron. Soc.},
  volume = {436},
  number = {4},
  pages = {3480--3491},
  year = {2013},
  month = dec,
  publisher = {Oxford University Press},
  doi = {10.1093/mnras/stt1836},
  bibcode = {2013MNRAS.436.3480P}
}

@article{Ceccarelli2013,
  author = {Ceccarelli, L. and others},
  title = {Clues on void evolution -- I. Large-scale galaxy distributions around voids},
  journal = {Mon. Not. Roy. Astron. Soc.},
  volume = {434},
  number = {2},
  pages = {1435--1442},
  year = {2013},
  month = sep,
  publisher = {Oxford University Press},
  doi = {10.1093/mnras/stt1097},
  bibcode = {2013MNRAS.434.1435C}
}

@article{Pisani2015,
  author = {Pisani, A. and others},
  title = {Counting voids to probe dark energy},
  journal = {Phys. Rev. D},
  volume = {92},
  number = {8},
  pages = {083531},
  year = {2015},
  doi = {10.1103/PhysRevD.92.083531}
}

@article{Bos2012,
  author = {Bos, E. G. Patrick and others},
  title = {The darkness that shaped the void: dark energy and cosmic voids},
  journal = {Mon. Not. Roy. Astron. Soc.},
  volume = {426},
  number = {1},
  pages = {440--461},
  year = {2012},
  doi = {10.1111/j.1365-2966.2012.21478.x},
  url = {https://academic.oup.com/mnras/article/426/1/440/1008768}
}

@article{Pourtsidou2026,
  author        = {Pourtsidou, Alkistis},
  title         = {Exponential quintessence with momentum coupling to dark matter},
  journal       = {JCAP},
  volume        = {2026},
  number        = {02},
  pages         = {014},
  year          = {2026},
  doi           = {10.1088/1475-7516/2026/02/014},
  eprint        = {2509.15091},
  archivePrefix = {arXiv},
  primaryClass  = {astro-ph.CO},
}

@article{Biswas2010,
  author = {Biswas, R. and Alizadeh, E. and Wandelt, B. D.},
  title = {Voids as a precision probe of dark energy},
  journal = {Phys. Rev. D},
  volume = {82},
  number = {2},
  pages = {023002},
  year = {2010},
  doi = {10.1103/PhysRevD.82.023002}
}

@article{Kreisch2019,
    author = {Kreisch, Christina D and others},
    title = {Massive neutrinos leave fingerprints on cosmic voids},
    journal = {Mon. Not. Roy. Astron. Soc.},
    volume = {488},
    number = {3},
    pages = {4413--4426},
    year = {2019},
    month = {09},
    doi = {10.1093/mnras/stz1944},
    url = {https://doi.org/10.1093/mnras/stz1944}
}

@article{Massara2015,
    author = {Massara, Elena and others},
    title = {Voids in massive neutrino cosmologies},
    journal = {JCAP},
    volume = {2015},
    number = {11},
    pages = {018},
    year = {2015},
    month = {11},
    doi = {10.1088/1475-7516/2015/11/018},
    url = {https://doi.org/10.1088/1475-7516/2015/11/018}
}

@article{Banerjee2016,
    author = {Banerjee, Arka and Dalal, Neal},
    title = {Simulating nonlinear cosmological structure formation with massive neutrinos},
    journal = {JCAP},
    volume = {2016},
    number = {11},
    pages = {015},
    year = {2016},
    month = {11},
    doi = {10.1088/1475-7516/2016/11/015},
    url = {https://doi.org/10.1088/1475-7516/2016/11/015}
}

@article{Schuster2019,
    author = {Schuster, Nico and others},
    title = {The bias of cosmic voids in the presence of massive neutrinos},
    journal = {JCAP},
    volume = {2019},
    number = {12},
    pages = {055},
    year = {2019},
    month = {12},
    doi = {10.1088/1475-7516/2019/12/055},
    url = {https://doi.org/10.1088/1475-7516/2019/12/055}
}

@article{Zhang2020,
    author = {Zhang, Gemma and others},
    title = {Void halo mass function: A promising probe of neutrino mass},
    journal = {Phys. Rev. D},
    volume = {102},
    number = {8},
    pages = {083537},
    year = {2020},
    month = {10},
    doi = {10.1103/PhysRevD.102.083537},
    url = {https://doi.org/10.1103/PhysRevD.102.083537}
}

@article{Contarini2021,
    author = {Contarini, Sofia and others},
    title = {Cosmic voids in modified gravity models with massive neutrinos},
    journal = {Mon. Not. Roy. Astron. Soc.},
    volume = {504},
    number = {4},
    pages = {5021--5038},
    year = {2021},
    month = {07},
    doi = {10.1093/mnras/stab1112},
    url = {https://doi.org/10.1093/mnras/stab1112}
}

@article{Bayer2021,
    author = {Bayer, Adrian E. and others},
    title = {Detecting Neutrino Mass by Combining Matter Clustering, Halos, and Voids},
    journal = {Astrophys. J.},
    volume = {919},
    number = {1},
    pages = {24},
    year = {2021},
    month = {09},
    doi = {10.3847/1538-4357/ac0e91},
    url = {https://doi.org/10.3847/1538-4357/ac0e91}
}

@article{Kreisch2022,
    author = {Kreisch, Christina D. and others},
    title = {The GIGANTES Data Set: Precision Cosmology from Voids in the Machine-learning Era},
    journal = {Astrophys. J.},
    volume = {935},
    number = {2},
    pages = {100},
    year = {2022},
    month = {08},
    doi = {10.3847/1538-4357/ac7d4b},
    url = {https://doi.org/10.3847/1538-4357/ac7d4b}
}

@article{Zivick2015,
    author = {Zivick, Paul and others},
    title = {Using cosmic voids to distinguish f(R) gravity in future galaxy surveys},
    journal = {Mon. Not. Roy. Astron. Soc.},
    volume = {451},
    number = {4},
    pages = {4215--4222},
    year = {2015},
    doi = {10.1093/mnras/stv1209},
    url = {https://doi.org/10.1093/mnras/stv1209}
}

@article{Barreira2015,
    author = {Barreira, Alexandre and others},
    title = {Weak lensing by voids in modified lensing potentials},
    journal = {JCAP},
    volume = {2015},
    number = {08},
    pages = {028},
    year = {2015},
    doi = {10.1088/1475-7516/2015/08/028},
    url = {https://doi.org/10.1088/1475-7516/2015/08/028}
}

@article{Falck2018,
    author = {Falck, Bridget and others},
    title = {Using voids to unscreen modified gravity},
    journal = {Mon. Not. Roy. Astron. Soc.},
    volume = {475},
    number = {3},
    pages = {3262--3272},
    year = {2018},
    month = {04},
    doi = {10.1093/mnras/stx3288},
    url = {https://doi.org/10.1093/mnras/stx3288}
}

@article{Baker2018,
    author = {Baker, Tessa},
    title = {Void lensing as a test of gravity},
    journal = {Phys. Rev. D},
    volume = {98},
    number = {2},
    pages = {023511},
    year = {2018},
    doi = {10.1103/PhysRevD.98.023511},
    url = {https://doi.org/10.1103/PhysRevD.98.023511}
}

@article{Paillas2019,
    author = {Paillas, Enrique and others},
    title = {The Santiago--Harvard--Edinburgh--Durham void comparison II: unveiling the Vainshtein screening using weak lensing},
    journal = {Mon. Not. Roy. Astron. Soc.},
    volume = {484},
    number = {1},
    pages = {1149--1165},
    year = {2019},
    month = {03},
    doi = {10.1093/mnras/stz022},
    url = {https://doi.org/10.1093/mnras/stz022},
    note = {Published: 05 January 2019}
}

@article{Davies2019,
    author = {Davies, Christopher T and Cautun, Marius and Li, Baojiu},
    title = {Cosmological test of gravity using weak lensing voids},
    journal = {Mon. Not. Roy. Astron. Soc.},
    volume = {490},
    number = {4},
    pages = {4907--4917},
    year = {2019},
    doi = {10.1093/mnras/stz2933},
    url = {https://doi.org/10.1093/mnras/stz2933}
}

@article{Perico2019,
    author = {Perico, Eder L. D. and others},
    title = {Cosmic voids in modified gravity scenarios},
    journal = {Astro. Astrophys.},
    volume = {632},
    pages = {A52},
    year = {2019},
    doi = {10.1051/0004-6361/201935949},
    url = {https://doi.org/10.1051/0004-6361/201935949}
}

@article{Tamosiunas2022,
    author = {Tamosiunas, Andrius and others},
    title = {Chameleon screening in cosmic voids},
    journal = {JCAP},
    volume = {2022},
    number = {11},
    pages = {056},
    year = {2022},
    doi = {10.1088/1475-7516/2022/11/056},
    url = {https://doi.org/10.1088/1475-7516/2022/11/056}
}

@article{Fiorini2022,
    author = {Fiorini, Bartolomeo and others},
    title = {Studying large-scale structure probes of modified gravity with COLA},
    journal = {JCAP},
    volume = {2022},
    number = {12},
    pages = {028},
    year = {2022},
    doi = {10.1088/1475-7516/2022/12/028},
    url = {https://doi.org/10.1088/1475-7516/2022/12/028}
}

@article{Lee2009,
    author = {Lee, Jounghun and Park, Daeseong},
    title = {Constraining the Dark Energy Equation of State with Cosmic Voids},
    journal = {Astrophys. J.},
    volume = {696},
    number = {1},
    pages = {L10},
    year = {2009},
    month = {04},
    doi = {10.1088/0004-637X/696/1/L10},
    url = {https://doi.org/10.1088/0004-637X/696/1/L10}
}

@article{Lavaux2012,
    author = {Lavaux, Guilhem and Wandelt, Benjamin D.},
    title = {Precision Cosmography with Stacked Voids},
    journal = {Astrophys. J.},
    volume = {754},
    number = {2},
    pages = {109},
    year = {2012},
    month = {07},
    doi = {10.1088/0004-637X/754/2/109},
    url = {https://doi.org/10.1088/0004-637X/754/2/109}
}

@article{Sutter2015b,
    author = {Sutter, P. M. and others},
    title = {On the observability of coupled dark energy with cosmic voids},
    journal = {Mon. Not. Roy. Astron. Soc. Lett.},
    volume = {446},
    number = {1},
    pages = {L1--L5},
    year = {2015},
    month = {01},
    doi = {10.1093/mnrasl/slu155},
    url = {https://doi.org/10.1093/mnrasl/slu155}
}

@article{Pollina2016,
    author = {Pollina, Giorgia and others},
    title = {Cosmic voids in coupled dark energy cosmologies: the impact of halo bias},
    journal = {Mon. Not. Roy. Astron. Soc.},
    volume = {455},
    number = {3},
    pages = {3075--3085},
    year = {2016},
    doi = {10.1093/mnras/stv2503},
    url = {https://doi.org/10.1093/mnras/stv2503}
}

@article{Verza2019,
    author = {Verza, Giovanni and others},
    title = {The void size function in dynamical dark energy cosmologies},
    journal = {JCAP},
    volume = {2019},
    number = {12},
    pages = {040},
    year = {2019},
    doi = {10.1088/1475-7516/2019/12/040},
    url = {https://doi.org/10.1088/1475-7516/2019/12/040}
}

@article{Verza2023,
    author = {Verza, Giovanni and others},
    title = {DEMNUni: disentangling dark energy from massive neutrinos with the void size function},
    journal = {JCAP},
    volume = {2023},
    number = {12},
    pages = {044},
    year = {2023},
    doi = {10.1088/1475-7516/2023/12/044},
    url = {https://doi.org/10.1088/1475-7516/2023/12/044}
}

@article{Hamaus2014b,
    author = {Hamaus, Nico and others},
    title = {Testing cosmic geometry without dynamic distortions using voids},
    journal = {JCAP},
    volume = {2014},
    number = {12},
    pages = {013},
    year = {2014},
    doi = {10.1088/1475-7516/2014/12/013},
    url = {https://doi.org/10.1088/1475-7516/2014/12/013}
}

@article{Hamaus2015,
    author = {Hamaus, Nico and others},
    title = {Probing cosmology and gravity with redshift-space distortions around voids},
    journal = {JCAP},
    volume = {2015},
    number = {11},
    pages = {036},
    year = {2015},
    doi = {10.1088/1475-7516/2015/11/036},
    url = {https://doi.org/10.1088/1475-7516/2015/11/036}
}

@article{Hamaus2016,
    author = {Hamaus, Nico and others},
    title = {Constraints on Cosmology and Gravity from the Dynamics of Voids},
    journal = {Phys. Rev. Lett.},
    volume = {117},
    number = {9},
    pages = {091302},
    year = {2016},
    doi = {10.1103/PhysRevLett.117.091302},
    url = {https://doi.org/10.1103/PhysRevLett.117.091302}
}

@article{Hawken2017,
    author = {Hawken, A. J. and others},
    title = {The VIMOS Public Extragalactic Redshift Survey: Measuring the growth rate of structure around cosmic voids},
    journal = {Astron.  Astrophys.},
    volume = {607},
    pages = {A54},
    year = {2017},
    doi = {10.1051/0004-6361/201629678},
    url = {https://doi.org/10.1051/0004-6361/201629678}
}

@article{Hamaus2017,
    author = {Hamaus, Nico and others},
    title = {Multipole analysis of redshift-space distortions around cosmic voids},
    journal = {JCAP},
    volume = {2017},
    number = {07},
    pages = {014},
    year = {2017},
    doi = {10.1088/1475-7516/2017/07/014},
    url = {https://doi.org/10.1088/1475-7516/2017/07/014}
}

@article{Massara2018,
    author = {Massara, Elena and Sheth, Ravi K.},
    title = {Density and velocity profiles around cosmic voids},
    journal = {arXiv e-prints},
    eprint = {1811.03132},
    archivePrefix = {arXiv},
    year = {2018},
    doi = {10.48550/arXiv.1811.03132},
    url = {https://doi.org/10.48550/arXiv.1811.03132}
}

@article{Correa2022,
    author = {Correa, Carlos M and others},
    title = {Redshift-space effects in voids and their impact on cosmological tests -- II. The void-galaxy cross-correlation function},
    journal = {Mon. Not. Roy. Astron. Soc.},
    volume = {509},
    number = {2},
    pages = {1871--1884},
    year = {2022},
    doi = {10.1093/mnras/stab3070},
    url = {https://doi.org/10.1093/mnras/stab3070}
}

@article{Massara2022,
    author = {Massara, Elena and others},
    title = {Velocity profiles of matter and biased tracers around voids},
    journal = {Mon. Not. Roy. Astron. Soc.},
    volume = {517},
    number = {3},
    pages = {4458--4471},
    year = {2022},
    doi = {10.1093/mnras/stac2892},
    url = {https://doi.org/10.1093/mnras/stac2892}
}

@article{Nadathur2019,
    author = {Nadathur, Seshadri and Percival, Will J.},
    title = {An accurate linear model for redshift space distortions in the void--galaxy correlation function},
    journal = {Mon. Not. Roy. Astron. Soc.},
    volume = {483},
    number = {3},
    pages = {3472--3487},
    year = {2019},
    doi = {10.1093/mnras/sty3372},
    url = {https://doi.org/10.1093/mnras/sty3372}
}

@article{Aghanim2020b,
    author = {Aghanim, N. and others},
    title = {Planck 2018 results: I. Overview and the cosmological legacy of Planck},
    journal = {Astron. Astrophys.},
    volume = {641},
    pages = {A1},
    year = {2020},
    doi = {10.1051/0004-6361/201833880},
    url = {https://doi.org/10.1051/0004-6361/201833880}
}

@article{Zeldovich1970,
    author = {Zel'dovich, Ya. B.},
    title = {Gravitational instability: An approximate theory for large density perturbations},

    journal = {Astron. Astrophys.},
    volume = {5},
    pages = {84--89},
    year = {1970}
}

@article{Gregory1978,
    author = {Gregory, S. A. and Thompson, L. A.},
    title = {The Coma/A1367 supercluster and its environs},
    journal = {Astrophys. J.},
    volume = {222},
    pages = {784--799},

    year = {1978},
    month = {06},
    doi = {10.1086/156198},
    url = {https://doi.org/10.1086/156198}
}

@article{Hamaus2022,
  author = {Hamaus, Nico. and others},
  title = {Euclid: Forecasts from redshift-space distortions and the Alcock--Paczynski test with cosmic voids},

  journal = {Astro. Astrophys.},
  volume = {658},
  pages = {A20},
  year = {2022},
  month = feb,
  doi = {10.1051/0004-6361/202142073},
  url = {https://doi.org/10.1051/0004-6361/202142073}
}

@article{Scaramella2022,
    author = {Scaramella, R. and others},
    title = {Euclid preparation. I. The Euclid Wide Survey},
    journal = {Astron. Astrophys.},
    volume = {662},
    pages = {A112},
    year = {2022},
    doi = {10.1051/0004-6361/202141938},
    url = {https://doi.org/10.1051/0004-6361/202141938},
    eprint = {2108.01201}
}

@article{Cropper2025,
    author = {Cropper, M. S. and others},
    title = {Euclid: II. The VIS instrument},
    journal = {Astron. Astrophys.},
    volume = {697},
    pages = {A2},
    year = {2025},
    doi = {10.1051/0004-6361/202450996},
    url = {https://doi.org/10.1051/0004-6361/202450996},
    eprint = {2405.13492}
}

@article{Spergel2015,
    author = {Spergel, D. and others},

    title = {Wide-Field InfrarRed Survey Telescope-Astrophysics Focused Telescope Assets WFIRST-AFTA 2015 Report},
    journal = {arXiv e-prints},
    eprint = {1503.03757},
    archivePrefix = {arXiv},
    year = {2015}
}

@article{AbdulKarim2025,
  author = {Abdul Karim, M. and Aguilar, J. and Ahlen, S. and others},
  title = {DESI DR2 results. II. Measurements of baryon acoustic oscillations and cosmological constraints},
  journal = {Phys. Rev. D},
  volume = {112},
  pages = {083515},
  year = {2025},
  doi = {10.1103/PhysRevD.112.083515}
}

@article{DESCollaboration2024,
  author = {Abbott, T. M. C. and others},
  title = {The Dark Energy Survey: Cosmology Results with ∼1500 New High-redshift Type Ia Supernovae Using the Full 5 yr Data Set},
  journal = {Astrophys. J. Lett.},
  volume = {973},
  number = {1},
  pages = {L14},
  year = {2024},
  doi = {10.3847/2041-8213/ad6f9f}
}

@article{Torrado2021,
  author = {Torrado, Jesús and Lewis, Antony},
  title = {Cobaya: code for Bayesian analysis of hierarchical physical models},
  journal = {JCAP},
  volume = {2021},
  number = {05},
  pages = {057},
  year = {2021},
  doi = {10.1088/1475-7516/2021/05/057}
}

@article{Gelman1992,
  author = {Gelman, Andrew and Rubin, Donald B.},
  title = {Inference from Iterative Simulation Using Multiple Sequences},
  journal = {Statistical Science},
  volume = {7},
  number = {4},
  pages = {457--472},
  year = {1992},
  doi = {10.1214/ss/1177011136}
}

@article{Lewis2025,
  author = {Lewis, Antony},
  title = {GetDist: a Python package for analysing Monte Carlo samples},
  journal = {Journal of Cosmology and Astroparticle Physics},
  volume = {2025},
  number = {08},
  pages = {025},
  year = {2025},
  doi = {10.1088/1475-7516/2025/08/025}}

@article{planck2020b,
  author = {Aghanim (Planck), N. and others},
  title = {Planck 2018 results. V. CMB power spectra and likelihoods},
  journal = {Astron. Astrophys.},
  volume = {641},
  pages = {A5},
  year = {2020},
  doi = {10.1051/0004-6361/201936386},
  archivePrefix = {arXiv},
  eprint = {1907.12875},
  bibcode = {2020A&A...641A...5P}
}

@article{Rosenberg2022,
  author = {Rosenberg, Erik and Gratton, Steven and Efstathiou, George},
  title = {CMB power spectra and cosmological parameters from Planck PR4 with CamSpec},
  journal = {Mon. Not. Roy. Astron. Soc.},
  volume = {517},
  number = {3},
  pages = {4620--4636},
  year = {2022},
  doi = {10.1093/mnras/stac2744}
}

@article{Carron2022,
  author = {Carron, Julien and Mirmelstein, Mark and Lewis, Antony},
  title = {CMB lensing from Planck PR4 maps},
  journal = {JCAP},
  volume = {2022},
  number = {09},
  pages = {039},
  year = {2022},
  doi = {10.1088/1475-7516/2022/09/039}
}

@book{baumann2022,
  author    = {Baumann, D.},
  title     = {Cosmology},
  chapter   = {2-6},
  publisher = {Cambridge University Press},
  year      = 2022,
  doi       = {10.1017/9781108937092}
}

@article{di_valentino2020b,
  author    = {Di Valentino, E. and others},
  title     = {Nonminimal Dark Sector Physics and Cosmological Tensions},
  journal   = {Phys. Rev. D},
  volume    = {101},
  number    = {6},
  pages     = {063502},
  year      = {2020},
  doi       = {10.1103/physrevd.101.063502}
}

@article{di_valentino2021,
  author    = {Di Valentino, E. and others},
  title     = {Cosmology Intertwined III: $f\sigma_8$ and $S_8$},
  journal   = {Astropart. Phys.},
  volume    = {131},
  pages     = {102604},
  year      = {2021},
  doi       = {10.1016/j.astropartphys.2021.102604}
}

@article{amendola2000,
  author    = {Amendola, L.},
  title     = {Coupled Quintessence},
  journal   = {Phys. Rev. D},
  volume    = {62},
  number    = {4},
  pages     = {043511},
  year      = {2000},
  doi       = {10.1103/physrevd.62.043511}
}

@article{pourtsidou2013,
  author    = {Pourtsidou, A. and Skordis, C. and Copeland, E.J.},
  title     = {Models of Dark Matter Coupled to Dark Energy},
  journal   = {Phys. Rev. D},
  volume    = {88},
  number    = {8},
  pages     = {083505},
  year      = {2013},
  doi       = {10.1103/physrevd.88.083505}
}

@article{pourtsidou2016,
  author    = {Pourtsidou, A. and Tram, T.},
  title     = {Reconciling CMB and Structure Growth Measurements with Dark Energy Interactions},
  journal   = {Phys. Rev. D},
  volume    = {94},
  number    = {4},
  pages     = {043518},
  year      = {2016},
  doi       = {10.1103/physrevd.94.043518}
}

@article{di_valentino2021b,
  author    = {Di Valentino, E. and Mukherjee, A. and Sen, A.A.},
  title     = {Dark Energy with Phantom Crossing and the H0 Tension},
  journal   = {Entropy},
  volume    = {23},
  number    = {4},
  pages     = {404},
  year      = {2021},
  doi       = {10.3390/e23040404}
}

@article{damico2016,
  author    = {D’Amico, G. and Hamill, T. and Kaloper, N.},
  title     = {Quantum Field Theory of Interacting Dark Matter and Dark Energy: Dark Monodromies},
  journal   = {Phys. Rev. D},
  volume    = {94},
  number    = {10},
  pages     = {103526},
  year      = {2016},
  doi       = {10.1103/physrevd.94.103526}
}

@article{marsh2017,
  author    = {Marsh, M. C. D.},
  title     = {Exacerbating the Cosmological Constant Problem with Interacting Dark Energy Models},
  journal   = {Phys. Rev. Lett.},
  volume    = {118},
  number    = {1},
  pages     = {011302},
  year      = {2017},
  doi       = {10.1103/physrevlett.118.011302}
}

@article{valiviita2008,
  author    = {Väliviita, J. and Majerotto, E. and Maartens, R.},
  title     = {Large-Scale Instability in Interacting Dark Energy and Dark Matter Fluids},
  journal   = {JCAP},
  volume    = {2008},
  number    = {07},
  pages     = {020},
  year      = {2008},
  doi       = {10.1088/1475-7516/2008/07/020}
}

@article{chamings2020,
  author    = {Chamings, F.N. and others},
  title     = {Understanding the Suppression of Structure Formation from Dark Matter-Dark Energy Momentum Coupling},
  journal   = {Phys. Rev. D},
  volume    = {101},
  number    = {4},
  pages      = {043531},
  year      = {2020},
  doi       = {10.1103/physrevd.101.043531}
}

@article{lesgourgues2011,
  author    = {Lesgourgues, J.},
  title     = {The Cosmic Linear Anisotropy Solving System (CLASS) I: Overview},
  journal   = {arXiv.org},
  year      = {2011},
  url       = {https://arxiv.org/abs/1104.2932},
}

@article{zhang2024,
  author    = {Zhang, W. and {Chu}, Ming-chung and Hu, R. and Liao, S. and Shek, Y.},
  title     = {Measuring Neutrino Mass and Asymmetry with Matter Pairwise Velocities},
  journal   = {Mon. Not. Roy. Astron. Soc.},
  volume    = {529},
  number    = {1},
  pages     = {360--373},
  year      = {2024},
  doi       = {10.1093/mnras/stae511}
}

@article{Aghanim2019,
  author = {Aghanim (Planck), N. and others},
  title = {Planck 2018 results. {VIII}. Gravitational lensing},
  journal = {Astro. Astrophys.},
  volume = {641},
  pages = {A8},
  year = {2019},
  doi = {10.1051/0004-6361/201833886},
  archiveprefix = {arXiv}
}

@article{Springel2005,
  title = {The Cosmological Simulation Code {GADGET-2}},
  author = {Springel, Volker},
  journal = {Mon. Not. Roy. Astron. Soc.},
  volume = {364},
  number = {4},
  pages = {1105--1134},
  year = {2005},
  doi = {10.1111/j.1365-2966.2005.09655.x},
  url = {https://doi.org/10.1111/j.1365-2966.2005.09655.x}
}

@article{An2019a,
  title = {Testing a Quintessence Model with Yukawa Interaction from Cosmological Observations and N-body Simulations},
  author = {An, Rui and others},
  journal = {Mon. Not. Roy. Astron. Soc.},
  volume = {489},
  number = {1},
  pages = {297--309},
  year = {2019},
  doi = {10.1093/mnras/stz2028},
  url = {https://doi.org/10.1093/mnras/stz2028}
}

@article{An2019b,
  title = {The First Constraint from {SDSS} Galaxy–Galaxy Weak Lensing Measurements on Interacting Dark Energy Models},
  author = {An, Rui and others},
  journal = {Astrophys. J.},
  volume = {875},
  number = {2},
  pages = {L11},
  year = {2019},
  doi = {10.3847/2041-8213/ab133f},
  url = {https://doi.org/10.3847/2041-8213/ab133f}
}

@article{zhang_gadget_2018,
  title = {Fully Self-Consistent Cosmological Simulation Pipeline for Interacting Dark Energy Models},
  author = {Zhang, Jiajun and others},
  journal = {Phys. Rev. D},
  volume = {98},
  number = {10},
  pages = {103530},
  year = {2018},
  doi = {10.1103/physrevd.98.103530},
  url = {https://doi.org/10.1103/physrevd.98.103530}
}

@article{Behroozi2012,
  title = {The {ROCKSTAR} Phase-Space Temporal Halo Finder and the Velocity Offsets of Cluster Cores},
  author = {Behroozi, Peter S. and Wechsler, Risa H. and Wu, Hao-Yi},
  journal = {Astrophys. J.},
  volume = {762},
  number = {2},
  pages = {109},
  year = {2012},
  doi = {10.1088/0004-637X/762/2/109},
  url = {https://doi.org/10.1088/0004-637X/762/2/109}
}

@article{Adame2025a,
  author = {Adame, A. G. and others},
  title = {{DESI 2024 III: Baryon Acoustic Oscillations from Galaxies and Quasars}},
  journal = {JCAP},
  volume = {2025},
  number = {04},
  pages = {012},
  year = {2025},
  doi = {10.1088/1475-7516/2025/04/012}
}

@article{Adame2025b,
  author = {Adame, A. G. and others},
  title = {{DESI 2024 VI: Cosmological Constraints from the Measurements of Baryon Acoustic Oscillations}},
  journal = {JCAP},
  volume = {2025},
  number = {02},
  pages = {021},
  year = {2025},
  doi = {10.1088/1475-7516/2025/02/021}
}

@article{Adame2025c,
  author = {Adame, A. G. and others},
  title = {{DESI 2024 IV: Baryon Acoustic Oscillations from the Lyman Alpha Forest}},
  journal = {JCAP},
  volume = {2025},
  number = {01},
  pages = {124},
  year = {2025},
  doi = {10.1088/1475-7516/2025/01/124}
}

@article{GomezValent2020,
  author = {Gómez-Valent, Adrià and Pettorino, V. and Amendola, L.},
  title = {{Update on Coupled Dark Energy and the H0 Tension}},
  journal = {Phys. Rev. D},
  volume = {101},
  number = {12},
  pages = {123513},
  year = {2020},
  doi = {10.1103/PhysRevD.101.123513}
}

@article{Bean2008,
  author = {Bean, R. and Flanagan, {{\'E}. {\'E}.}. and Laszlo, I. and Trodden, M.},
  title = {{Constraining Interactions in Cosmology’s Dark Sector}},
  journal = {Phys. Rev. D},
  volume = {78},
  number = {12},
  pages = {123514},
  year = {2008},
  month = {Dec},
  publisher = {American Physical Society},
  doi = {10.1103/PhysRevD.78.123514}
}

@article{Xia2009,
  author = {Xia, J.-Q.},
  title = {{Constraint on Coupled Dark Energy Models from Observations}},
  journal = {Phys. Rev. D},
  volume = {80},
  number = {10},
  pages = {103514},
  year = {2009},
  doi = {10.1103/PhysRevD.80.103514}
}

\appendix*
\section{} \label{sec: appendix}

\setcounter{table}{0}
\renewcommand{\thetable}{A.\Roman{table}}
\renewcommand{\theHtable}{A.\Roman{table}}

\setcounter{figure}{0}                       
\renewcommand\thefigure{A.\arabic{figure}}
\renewcommand\theHfigure{A.\arabic{figure}}



\renewcommand{\thesubsection}{\Roman{subsection}}

\subsection{MCMC constraints}
\label{sec:mcmc_constraints}

Table~\ref{tab:cosmo_params} shows the cosmological parameters for the Type 3 model obtained in \cite{Pourtsidou2026}, with the prior ranges $\beta \in [-2.0,0.5]$ and $\lambda \in [0,2.1]$. Parameter values of $\beta$ and $\lambda$ within the $1\sigma$ lower and upper bounds of the constraints are used to perform the N-body simulations, as detailed in Table~\ref{tab:mcmc_cases}.

\begin{table}
\vspace{3mm}
    \centering
    \begin{tabular}{cc}
    \hline
    \hline
    Parameter & $\text{T3}$ \\
    \hline     
    $100\,\Omega_b h^2$ & $2.23 \pm 0.01$ \\
    $\Omega_{c} h^2$ & $0.117 \pm 0.001$ \\
    $10^4\theta_s$ & $104.20 \pm 0.02$ \\
    $\tau_{reio}$ & $0.06 \pm 0.01$ \\
    $n_s$ & $0.969 \pm 0.004$ \\
    $\sigma_8$ & $0.777^{+0.026}_{-0.015}$ \\
    $H_0$ & $66.9 \pm 0.6$ \\
    $\lambda$ & $1.00 \pm 0.4$ \\
    $\beta$ & $-0.8^{+1.0}_{-0.6}$ \\
    
    \hline
    \end{tabular}
    \caption{Best-fit values (68\% CL) of the cosmological parameters for the Type 3 model \cite{Pourtsidou2026}, obtained by fitting with the Planck 2018 CMB, lensing \cite{planck2020b, Rosenberg2022, Carron2022, Aghanim2019}, DESI DR2 BAO \cite{AbdulKarim2025}, and the DES-Y5 Type Ia supernova sample data \cite{DESCollaboration2024}. Parameter values corresponding to negative $\beta$ are used to perform the N-body simulations, as shown in Table~\ref{tab:mcmc_cases}.}
    \label{tab:cosmo_params}
\end{table}

\subsection{Void radial velocity statistics for CDM particle voids}

\label{sec:CDM_voids}
This section presents the void radial velocity statistics traced by CDM particles. The results show that, independent of tracer type, variations in $\beta$ and $\lambda$  induce the same qualitative changes in both the radial velocity and velocity-dispersion profiles. This indicates that void radial velocity statistics provide a robust probe of the Type 3 model parameters.

\begin{figure*}
    \includegraphics[width=.80\textwidth]{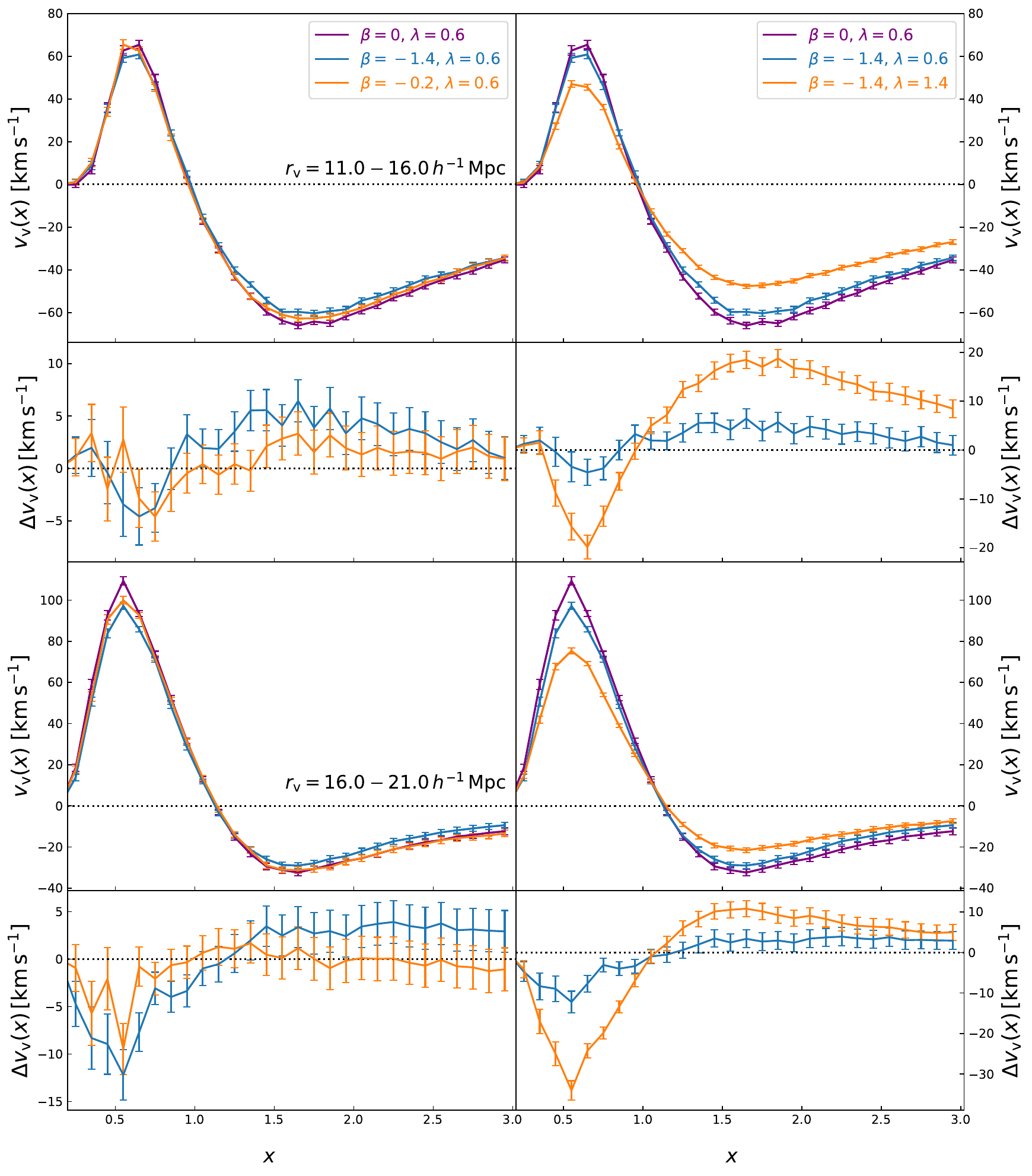}
    \caption{Same as Fig. \ref{fig:vel_h}, but for isolated CDM particle voids. For $r_\text{v}=11-16 \,h^{-1}\,\text{Mpc}$ and $r_\text{v}=16-21 \,h^{-1}\,\text{Mpc}$, each radial size bin contains approximately 9000 and 7000 voids, respectively.}\label{fig:vel_p}
\end{figure*}

\begin{figure*}
    \includegraphics[width=.80\textwidth]{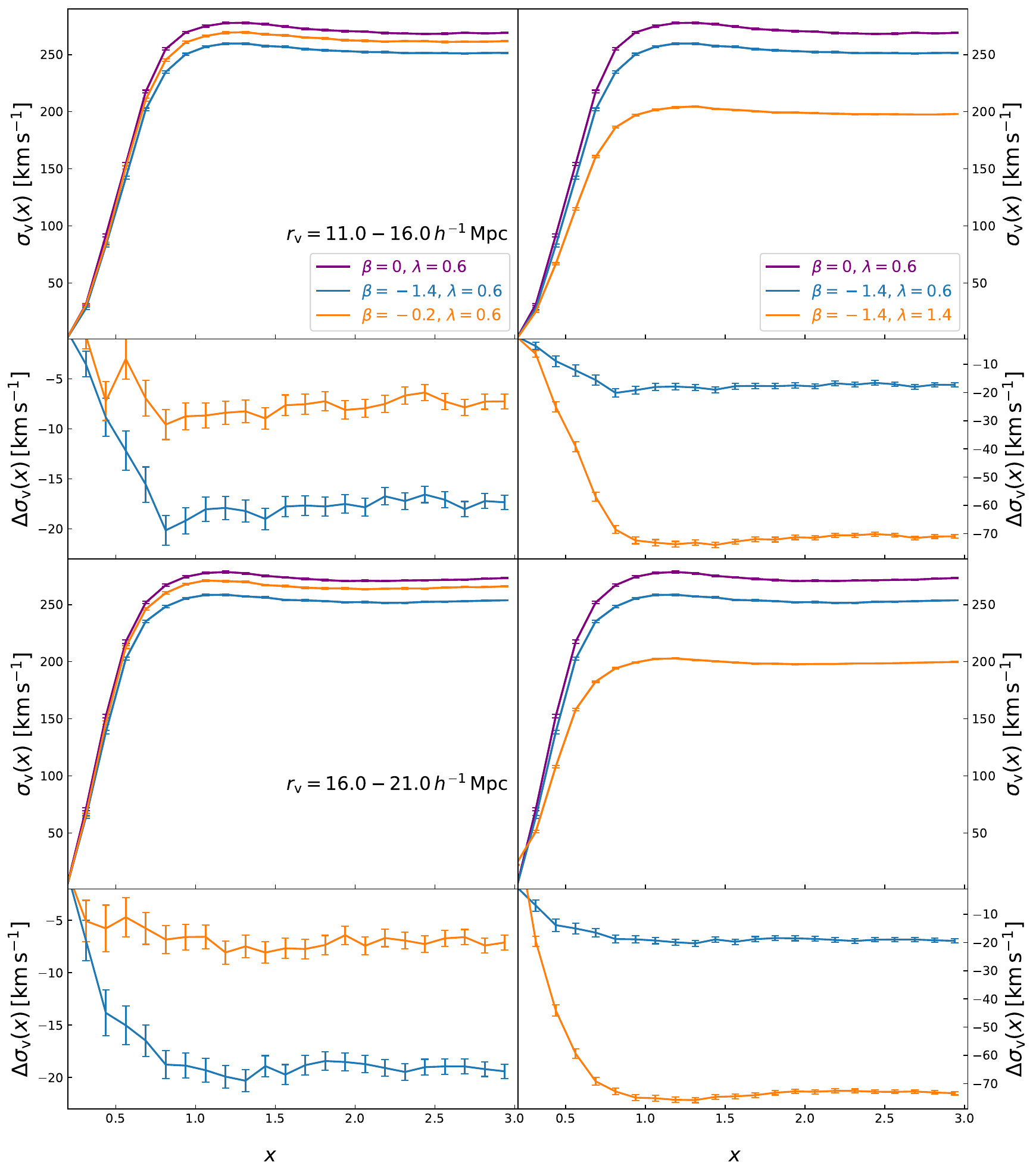}
    \caption{Same as Fig. \ref{fig:sigma_h}, but for isolated CDM particle voids. For $r_\text{v}=11-16 \,h^{-1}\,\text{Mpc}$ and $r_\text{v}=16-21 \,h^{-1}\,\text{Mpc}$, each radial size bin contains approximately 9000 and 7000 voids, respectively.}\label{fig:sigma_p}
\end{figure*}

\begin{table*}[b]
\centering
\renewcommand{\arraystretch}{1.3}
\begin{tabular}{l l c c c c c}
\hline\hline
Quantity & ${r_\text{v}}$ [$h^{-1}$\text{Mpc}] & $c_\beta \, (\%) $ & $c_\lambda \, (\%) $ & $c_{\beta\beta} \, (\%) $ & $c_{\beta\lambda} \, (\%) $ & $R^2$ $(\bar{R}^2)$ \\
\hline
\multirow{2}{*}{$\Delta \bar{v}_{\text{span}}$} 
& 11.0 -- 16.0 & $14.51 \pm 2.76$ & $-11.34 \pm 2.14$ & $7.13 \pm 2.13$ & $11.16 \pm 2.36$ & 0.987 (0.980) \\
& 16.0 -- 21.0 & $18.41 \pm 2.59$ & $-9.41 \pm 2.02$ & $9.28 \pm 2.00$ & $13.11 \pm 2.23$ & 0.992 (0.988) \\
\cline{1-7}
\multirow{2}{*}{$\Delta \bar{\sigma}_{\text{span}}$} 
& 11.0 -- 16.0 & $12.51 \pm 1.19$ & $-8.74 \pm 0.92$ & $6.39 \pm 0.91$ & $13.73 \pm 1.02$ & 0.996 (0.993)\\
& 16.0 -- 21.0 & $10.33 \pm 1.84$ & $-12.93 \pm 1.41$ & $4.78 \pm 1.41$ & $10.21 \pm 1.57$ & 0.990 (0.984)\\
\hline\hline
\end{tabular}
\caption{Best-fit coefficients for the regression of $\Delta \bar{v}_{\text{span}}$ and $\Delta \bar{\sigma}_{\text{span}}$ for CDM voids relative to the uncoupled A0 scenario. The final column provides the $R^2$ and $\bar{R}^2$ for each fit.}\label{coefficients_c}

\end{table*}

\end{document}